\documentstyle[prb,aps,multicol,epsf]{revtex}

\def\wid{\protect\end{multicols}
\widetext
\noindent\rule{20.5pc}{0.1mm}\rule{0.1mm}{1.5mm}\hfill}
\def\nar{\hfill\rule[-1.5mm]{0.1mm}{1.5mm}\rule{20.5pc}{0.1mm}
\begin{multicols}{2}
\narrowtext }
\newcommand{\bfpsi}{\mbox{\boldmath $\psi$}}
\newcommand{\bfphi}{\mbox{\boldmath $\phi$}}

\begin{document}

\draft

\title{Tunneling decay in a magnetic field}

\author{ T.~Sharpee$^{(a)}$, M.I.~Dykman$^{(a)}$\cite{byline}, 
P.M.~Platzman$^{(b)}$}
\address{$^{(a)}$Department of Physics and Astronomy, 
Michigan State University, East Lansing, Michigan 48824\\
$^{(b)}$Bell Laboratories, Lucent Technologies, Murray Hill, New Jersey 07974}
\date{\today} 
\maketitle
\widetext 
\begin{quote} 
We provide a semiclassical theory of tunneling decay in a magnetic
field and a three-dimensional potential of a general form. Because of
broken time-reversal symmetry, the standard WKB technique has to be
modified. The decay rate is found from the analysis of the set of the
particle Hamiltonian trajectories in complex phase space and time. In
a magnetic field, the tunneling particle comes out from the barrier
with a finite velocity and behind the boundary of the classically
allowed region. The exit location is obtained by matching the decaying
and outgoing WKB waves at a caustic in complex configuration
space. Different branches of the WKB wave function match on the {\it
switching} surface in real space, where the slope of the wave function
sharply changes. The theory is not limited to tunneling from potential
wells which are parabolic near the minimum. For parabolic wells, we
provide a bounce-type formulation in a magnetic field. The theory is
applied to specific models which are relevant to tunneling from
correlated two-dimensional electron systems in a magnetic field
parallel to the electron layer.
\end{quote} 

\pacs{PACS numbers: 03.65.Sq, 03.65.Xp, 73.40.Gk, 73.21.-b}

\begin{multicols}{2} 
\narrowtext 

\section{Introduction}
Magnetic field can very strongly change the tunneling rate of charged
particles. This change, in turn, strongly depends on properties of the
system, as in the well-known effect of giant hopping magnetoresistance
in solids \cite{Efros-Shklovskii}. Therefore tunneling in a magnetic
field has been broadly used as a sensitive and revealing probe of
electron systems in solids, including quantum Hall systems
\cite{DasSarma_book-97,Ashoori,Chang,Eisenstein-2000},
two-layer heterostructures away from the quantum Hall region
\cite{Smoliner,Eisenstein,MacDonald,minigaps}, and  
correlated electron layers on the surface of liquid helium
\cite{Andreimagn,Dykman-01}.

Correlated two-dimensional (2D) electron systems are currently
attracting much interest \cite{Abrahams-00}. The possibility to
extract information about electron correlations and dynamics through
tunneling in a magnetic field is one of the motivations of the present
work. The major motivation, however, comes from the fact that
tunneling in a magnetic field is an interesting and in many respects
unusual theoretical problem, even in the single-particle
formulation. Existing results, although often highly nontrivial, are
limited to the cases where the potential has either a special form
\cite{Popov,Caldeira,Fert_Halperin,Ao} (e.g., linear \cite{Popov} or parabolic
\cite{Fert_Halperin}), or a part of the potential or the magnetic
field are in some sense weak
\cite{Shklovskii,Thouless,Feng,Hajdu,Helffer,Nakamura}.

The problem of tunneling has two parts. One is to find the tail of the
wave function of the localized intrawell state under the potential
barrier $U({\bf r})$ and behind it, and the other is to find the
escape probability $W$. In the magnetic field, $W$ differs
exponentially from the probability to reach the boundary of the
classically allowed range $U({\bf r}) =E$, where $E$ is the energy of
the particle. This is because, as it tunnels, the particle is
accelerated by the Lorentz force, and it comes out from the barrier
with a finite velocity ${\bf v}$. The standard argument
that the exit point is the turning point ${\bf v=0}$ relies on
time-reversal symmetry (see below) and does not apply in the presence of
a magnetic field.

\begin{figure}
\begin{center}
\epsfxsize=2.4in                
\leavevmode\epsfbox{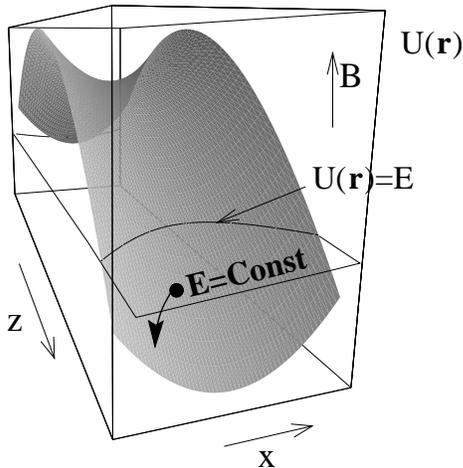}
\end{center}
\caption{Tunneling in a two-dimensional potential
$U(x,z)$ transverse to a magnetic field $B$ pointing in the $y$
direction. Initially the particle is localized in a metastable state
behind the barrier (on the small-$z$ side), with energy $E$. In
contrast to the case $B=0$, a particle emerges from the barrier with a
finite velocity, and therefore the exit point is located away from the
line $U({\bf r})=E.$}
\label{fig:3D-potential}
\end{figure}

\begin{figure}
\begin{center}
\epsfxsize=2.4in                
\leavevmode\epsfbox{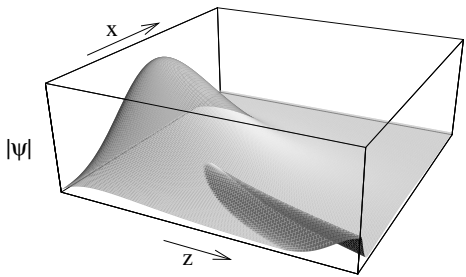}
\end{center}
\caption{The absolute value of the tunneling wave function (schematically). 
The maximum is located inside the potential well, i.e. behind the
barrier in Fig.~\protect\ref{fig:3D-potential}. A Gaussian wave packet
of nearly constant height describes the escaped particle, which shows
up with a finite velocity.}
\label{fig:wave-function}
\end{figure}

A simple potential $U({\bf r})$ and the wave function $\psi({\bf r})$
of the metastable state in this potential are sketched in
Figs.~\ref{fig:3D-potential},
\ref{fig:wave-function}.  The wave function decays away from the 
potential well. At some point ${\bf r}$, on the background of the
decaying tail there emerges a propagating small-amplitude wave packet,
which corresponds to the escaped particle.  As a result, in a part of
the classically allowed region $U({\bf r}) < E$ the function
$\psi({\bf r})$ is determined by this wave packet, whereas in the
other part of this region $\psi$ is determined by a different branch
of the tail of the localized state. The boundary between these areas
is sharp, and the slope of the wave function changes on this boundary
nearly discontinuously.

An important part of the WKB formulation of the tunneling escape
problem in a magnetic field was found \cite{Popov}$^{\rm (b)}$ in the
analysis of decay for a special model of an atomic system [see
Eq.~(\ref{exit}) below]. In a general case, both the tail of a
metastable state and the outgoing wave packet can be found using the
approach briefly outlined in our Letter
\cite{Barabash-2000}. 

In the WKB approximation the wave function is
sought in the form
\begin{equation}
\label{wavefunction}
\psi({\bf
r})=D({\bf r}) \exp [ iS({\bf r})] \quad (\hbar=1).
\end{equation}
Here, $S({\bf r})$ is the classical action, and $D$ is the prefactor.
In the classically allowed range, (\ref{wavefunction}) describes a
wave propagating with a real momentum ${\bf p} =
\bbox{\nabla} S$. On the other hand, in the classically forbidden
range the wave function decays. For the ground
intrawell state, the decay of $\psi({\bf r})$ is not
accompanied by oscillations in the absence of a magnetic field. Then
the action $S({\bf r})$ is purely imaginary under the barrier, and $
|\bbox{\nabla} S|$ is the decrement of the wave function.

Both behind and inside the barrier, the action can be obtained from
the Hamilton equations of motion
\begin{eqnarray}
\label{eom}
\dot S={\bf p}\cdot\dot{\bf r}, \quad \dot{\bf r} = \partial
H/\partial {\bf p},\;  \dot{\bf p} = -\partial
H/\partial {\bf r},
\end{eqnarray}
where $H$ is the Hamiltonian of the system,
\begin{equation}
\label{hamiltonian}
H={1\over 2m}\left[{\bf p}+{e\over c}{\bf
A}({\bf r})\right]^2 + U({\bf r}),
\end{equation} 
(${\bf A}({\bf r})$ is the vector potential of the magnetic field).

In the standard approach to tunneling decay, which applies for $B=0$
\cite{Caldeira,Langer,Coleman,AK,Schmid}, one looks for the purely 
imaginary action $S$ under the barrier. It is calculated by changing
to imaginary time and momentum in Eqs.~(\ref{eom}), but keeping
coordinates real,
\begin{equation}
\label{replacement}
t\to -it,\, {\bf p}
\to i{\bf p}, \, {\bf r}\to {\bf r}, \, U({\bf r}) \to -U({\bf r}).
\end{equation}
Eqs.(\ref{eom}) then take the form of equations of classical motion in
an inverted potential $-U({\bf r})$, with energy $-E\geq -U({\bf r})$.
The imaginary-time trajectory goes from the turning point inside the
potential well to the turning point on the boundary of the classically
allowed region, where it matches the appropriate classical trajectory
of the escaped particle behind the barrier, with real ${\bf
r,p},t$. The velocities on the both trajectories at the matching point
can coincide only if ${\bf v=0}$.

In the presence of a magnetic field, because of broken time-reversal
symmetry, the replacement (\ref{replacement}) may not be performed. It
would lead to a complex Hamiltonian, which makes no sense and
indicates that a more general approach is required.  The action
$S({\bf r})$ is complex under the barrier for real ${\bf r}$.  This
complexity has also an important counterpart in the instanton
formulation of the problem of tunneling decay in a magnetic field, see
below. We note that a complex action arises also in other cases, like
barrier penetration for oblique incidence
\cite{Huang} and scattering by a complex potential 
(as in the case of an absorbing medium) \cite{Knoll}. The method
discussed below can be applied to these problems as well.

In this paper we consider a single-particle tunneling decay in a
three-dimensional potential of a general form, for arbitrary magnetic
fields. We illustrate the approach using a toy model of a correlated
2D electron system. We show that the tunneling exponent $S({\bf r})$
and the escape rate in a magnetic field can be found from dynamical
equations (\ref{eom}) by analytically continuing these equations to
complex phase space and time. The initial conditions for the
trajectories are determined by the analytical continuation of the
usually known intrawell wave function. The resulting set of complex
trajectories has singularities, caustics, on which there occurs
branching of the complex action $S({\bf r})$ and the tails of the
decaying and propagating waves are matched. Careful analysis allows us
to find the complete semiclassical wave function and reveal the
singular features of $\psi({\bf r})$ related to the branching of
$S$. 

In Sec.~\ref{sec:model} and Appendix~\ref{app1} we provide a simple model
which catches basic physics of tunneling from correlated 2D electron
layers. In Sec.~\ref{sec:exp} we consider the tunneling exponent and
formulate the boundary value problem for tunneling trajectories in a
magnetic field in complex phase space.
In Sec.~\ref{sec:local} we discuss matching of different semiclassical
solutions across the caustic of the set of the tunneling
trajectories. We show that a switching surface (one of the anti-Stokes
manifolds) starts at the caustic. The wave function has an observable
singular feature at this surface, which is a sharp change of the slope
of $\ln|\psi({\bf r})|$. In Sec.~\ref{sec:correlated} we provide 
explicit results for two simple exactly solvable models of physical
interest, which also illustrate general features of tunneling in a
magnetic field. In Sec.~\ref{sec:instanton} we discuss the
path-integral formulation of the problem of tunneling decay in a
magnetic field.
Sec.~\ref{sec:conc} contains concluding remarks.

\section{A model of the tunneling barrier for a correlated 2D electron system}
\label{sec:model}

One of the most interesting systems where tunneling in a magnetic
field has been investigated experimentally is a correlated 2D electron
system. Here, electrons are localized in the $z$-direction in a
metastable 1D potential well $U_0(z)$. The intrawell electron motion
is quantized in the $z$-direction, and electrons can tunnel from the
well into extended states. Many 2D systems of current interest are
strongly correlated: electrons are far away from each other, exchange
is weak, and there is at least short-range order in the
$(x,y)$-plane. The tunneling electron can be then identified and
``labeled''. Its tunneling motion is accompanied by the motion of
other electrons. The many-electron dynamics of a correlated system can
be described in terms of in-plane electron vibrations, and the
corresponding Hamiltonian is given in Appendix~\ref{app1} assuming that the
electrons form a Wigner crystal. Here we will made a further
simplification and think of an electron as tunneling in a static
potential created by all other electrons. As we showed earlier, this
is a good approximation for the tunneling problem\cite{Dykman-01}.

The static potential from surrounding electrons can be assumed to be
parabolic with respect to the in-plane coordinates $x,y$. If the
characteristic width $L$ of the tunneling barrier is less than the
interelectron distance, the overall potential is a sum of the
parabolic in-plane part and $U_0(z)$,
\begin{equation}
\label{U(z)}
U({\bf r})= {m\omega_0^2\over 2}(x^2+y^2) + U_0(z).
\end{equation}

The form of $U_0(z)$ depends on the system. Inside the well $U_0$ is
often singular, like in the case of electrons on helium where $U_0$
includes the image potential. The potential barrier itself can be
close to a square barrier, as in the case of unbiased semiconductor
heterostructures, or can be nearly linear, as in the presence of
strong enough bias voltage, with

\begin{equation}
\label{U_0(z)}
U_0(z)={\gamma^2\over
2m}\left(1-{z\over L}\right) \; 
\end{equation}
inside the barrier. The potential (\ref{U_0(z)}) describes, in
particular, the barrier for a correlated 2DES on a helium
surface. This system was experimentally investigated in
Ref. \cite{Andreimagn}, and showed an unexpected dependence of the
tunneling rate on $B$ and electron density that we recently addressed
\cite{Dykman-01,Barabash-2000}.

\section{The tunneling exponent}
\label{sec:exp}


For a smooth tunneling barrier $U({\bf r})$, the underbarrier wave
function $\psi({\bf r})$ (\ref{wavefunction}) can be obtained from the
tunneling trajectories (\ref{eom}). The initial conditions for these
trajectories are determined by the tail of the intrawell wave
function. They can be obtained even if the potential $U({\bf r})$ is
singular within the well, as in the case of an image potential or a
stepwise potential in a semiconductor heterostructure.

To obtain the initial conditions for the trajectories (\ref{eom}), we
can take a surface $\Sigma$ close to the well and yet in the range
where $U({\bf r})$ is already smooth. The wave function $\psi({\bf
r})$ on $\Sigma$ is presumably known from the solution of the
Schr\"odinger equation inside the well and is semiclassical. Only the
exponent of this wave function is needed to find the initial
conditions for (\ref{eom}), which take the form

\begin{equation}
\label{initial_general}
{\bf r}(0) = {\bf r}|_{\Sigma}, \quad {\bf p}(0) =
-i\left[{\bbox\nabla}\ln \psi({\bf r})\right]_{\Sigma},
\end{equation}
with the action $S(0)=-i\left[\ln\psi({\bf r})\right]_{\Sigma}$.
Only the lowest-order terms in $\hbar$ should be kept in the
expressions for ${\bf p}(0),\,S(0)$. The final result should be
independent of the choice of $\Sigma$.

The trajectories (\ref{eom}) with the initial
conditions (\ref{initial_general}) form a {\it two-parameter} set, in
the case of 3D tunneling. The two parameters are the initial
coordinates on the surface $\Sigma$. We can choose curvilinear
coordinates $(x_1,x_2,x_3)$ so that $x_3|_{\Sigma}=0$, and
respectively $x_3(0)=0$. The trajectories are then parametrized by
$x_1(0),\, x_2(0)$.

To illustrate these arguments we consider the initial conditions for
an electron with the potential (\ref{U(z)}), (\ref{U_0(z)}), which
tunnels from a 2D layer. Inside the metastable potential well the
electron motion separates into a quantized motion in the normal to the
layer $z$-direction and in-plane vibrations. To slightly simplify the
analysis, we will neglect the effect of a magnetic field on the
intrawell wave function, but not on the tail of $\psi({\bf r})$ deep
under the barrier where the effect will have accumulated. This is a
good approximation for not too strong fields provided the
characteristic intrawell localization length $1/\gamma$ is small
compared to the tunneling length $L$.

It is convenient then to choose the surface $\Sigma$ as a plane
$z=$~const close to the well, but behind the intrawell turning point.
We set $z = x_3=0$ on $\Sigma$ and choose $x_1=x, x_2=y$. If we
set the energy of the out-of-plane motion $E=0$ and assume that the
electron is in the ground intrawell state, we obtain from (\ref{U(z)})
\begin{eqnarray}
\label{initial}
&&z(0)=0,\; p_z(0)=i\gamma,\; S(0)=im\omega_0 [x^2(0)+y^2(0)]/2,\nonumber\\
&& p_x(0)=im\omega_0x(0),\; p_y(0)=im\omega_0y(0).
\end{eqnarray}

To find the trajectory which arrives at a given real ${\bf r}$ deep
under the barrier, it may be necessary, particularly for $B\neq 0$, to
start with a {\it complex} ${\bf r}(0)$. The corresponding values of
${\bf p}(0)$ can be found by analytically continuing the intrawell
wave function to complex ${\bf r}$. The whole trajectory will then go
in complex phase space (including the configurational space) and also
in complex time $t$. The energy of the trajectory is given by the
energy of the intrawell state from which the system tunnels. It
remains real.

The rate of tunneling decay is determined by ${\rm Im}$~$S$ at the
point where the particle emerges from the barrier as a semiclassical
wave packet. This wave packet propagates along a {\it real} classical
trajectory ${\bf r}_{\rm cl}(t)$, which is a real-time solution of the
Hamiltonian equations (\ref{eom}). The underbarrier trajectory for
tunneling escape should coalesce with this classical
trajectory. Therefore at some time $t$ it should have both real
coordinate and velocity, 
\begin{equation}
\label{exit}
{\rm Im}~{\bf r}(t) = {\rm Im}~{\bf p}(t) = 0.
\end{equation}
\noindent

\begin{figure}
\begin{center}
\epsfxsize=3.4in                
\leavevmode\epsfbox{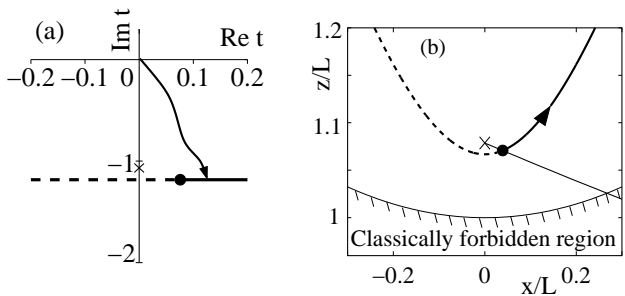}
\end{center}
\caption{(a) Complex $t$ plane for integrating the Hamiltonian
equations (\protect\ref{eom}) in the escape problem. The line Im~$t$ =
const corresponds to the classical trajectory of the escaped electron,
which is shown in (b). Bold solid lines in (a) and (b) indicate the
range where the amplitude of the propagating wave exceeds the
amplitude of the decaying underbarrier wave function. The escaped
particle shows up as a semiclassical wave packet, with a finite
velocity, at the point (full circle) where the classical trajectory
intersects the switching line [thin solid line in (b)]. The
crosses mark the value of $t$ (a) and the position (b) of the caustic
where it goes through real space. For the chosen parameter values, the
time when the escape trajectory hits the caustic is numerically very
close to the position of the cross in (a). The specific data refer to
tunneling through the potential barrier (\ref{U(z)}), (\ref{U_0(z)})
transverse to a magnetic field, which points in the $y$-direction,
with $\omega_0
\tau_0=1.2$ and $\omega_c\tau_0=1.2$; time in (a) is in the units of
$\tau_0=2mL/\gamma$. }
\label{fig:time}
\end{figure}

Eqs.~(\ref{exit}) determine the complex starting point of the
trajectory for tunneling escape ${\bf r}(0)$ [i.e., the complex
$x_{1,2}(0)$, since $x_3(0)=0$] and also the imaginary part of the
duration of motion along this trajectory Im~$t$. The real part of $t$
remains undetermined: a change in Re~$t$ in (\ref{exit}) results just
in a shift of the particle along the classical trajectory ${\bf
r}_{\rm cl}(t)$, see Fig.~\ref{fig:time}. Such a shift does not change
Im~$S$, since ${\bf p}= {\bbox\nabla}S$ is real along ${\bf r}_{\rm
cl}(t)$. We note that the number of equations (\ref{exit}) is equal to
the number of variables Re~$x_{1,2}(0)$, Im~$x_{1,2}(0)$, and Im~$t$
[the value $x_3(0)=0$ is fixed on $\Sigma$], with account taken of
energy conservation.  The conditions (\ref{exit}) were first given
\cite{Popov}$^{\rm (b)}$ for a $\delta$-shape potential well and a
linear tunneling barrier, but only the condition Im~${\bf r}(t)=0$ was
used.

In the absence of a magnetic field, we can choose the surface $\Sigma$
such that the momentum ${\bf p}|_{\Sigma}$ is imaginary for real ${\bf
r}$, i.e. the decay of the localized wave function is not accompanied
by oscillations [cf. (\ref{initial})]. The equations of motion
(\ref{eom}) can then be solved in purely imaginary time, with real
${\bf r}(t)$ and imaginary ${\bf p}(t)$. The escape trajectory ends at
the turning point ${\bf p}={\bf 0}$, even for a multidimensional
system \cite{Schmid}.

The tunneling exponent
$R$ is given by the value of Im~$S$ at any point on the
trajectory ${\bf r}_{\rm cl}$,
\begin{equation}
\label{answer}
R = 2~{\rm Im}~S({\bf r}_{\rm cl}).
\end{equation}
\noindent
For a physically meaningful solution, Im~$S$ should have a parabolic
minimum at ${\bf r}_{\rm cl}$ as a function of the coordinates transverse
to the trajectory. The outgoing wave packet will then be Gaussian
near the maximum.

From (\ref{exit}), even in the presence of a magnetic field the
tunneling exponent can be obtained by solving the equations of motion
(\ref{eom}) in imaginary time, with complex ${\bf r}$. However, such
solution does {\bf not} give the wave function for real ${\bf r}$
between the well and the classical trajectory ${\bf r}_{\rm cl}$. Neither
does it tell us {\bf where} the particle shows up on the classical
trajectory.

\section{Branching of the action and its observable consequences}
\label{sec:local}

The complete WKB solution of the tunneling problem can be obtained and
the wave function $\psi({\bf r})$ can be found if one takes into
account that the action $S$ as given by Eqs.~(\ref{eom}) is a
multivalued function of $\bf r$, even though it is a single-valued
function of $t$ and $x_{1,2}(0)$. This means that several trajectories
(\ref{eom}) with different $t$ and $x_{1,2}(0)$ can go through one and
the same point ${\bf r}$. However, except for the points on the
switching surface (see below), only one of the branches of the action
$S({\bf r})$ contributes to the wave function $\psi(\bf{r})$.

\subsection{Caustics in a magnetic field}

\begin{figure}[t]
\begin{center}
\epsfxsize=3.4in                
\leavevmode\epsfbox{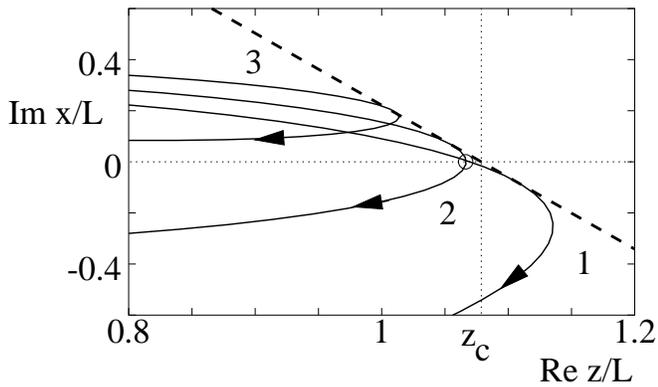}
\end{center}
\caption{The set of tunneling trajectories (solid
 lines) with Re~$x =$ Im~$z = 0$ and the caustic (dashed line) for an
effectively 2D tunneling problem. The data refer to the potential
(\ref{U(z)}), (\ref{U_0(z)}) and a magnetic field along the
$y$-axis. The parameters are the same as in Fig.~\ref{fig:time}. The
classical trajectory of the escaped electron is the real-time
continuation of the trajectory 2 from the point Im~$x=0$ shown by the
open circle. The momentum ${\bf p}$ is real at this point, as
explained in Sec.~\ref{subsec:symmetry}.  Other trajectories do not
describe escape, since for them the necessary conditions ${\rm
Im}~\dot z = {\rm Im}~x = 0$ are not satisfied. The caustic goes
through real space at the point $z=z_c,\, {\rm Im}~x=0$ (the
intersection of the dotted lines). The trajectory of the escaped
electron does {\it not} go through this point.}
\label{fig:caus}
\end{figure}

In multidimensional systems, branching of the semiclassical action
generally occurs on caustics, or envelopes of the Hamiltonian
trajectories \cite{Berry,Schulman_book}, see
Fig.~\ref{fig:caus}. Caustics are multidimensional counterparts of
turning points familiar from the analysis of tunneling in 1D systems
\cite{LandauQM}. The prefactor $D({\bf r})$ in the WKB wave function
(\ref{wavefunction}) diverges at a caustic. In the case of 1D
semiclassical motion along the $z$-axis we have $D\propto
p_z^{-1/2}$, and $D\to \infty$ at the turning points $z_t$, which are
given by the condition $p_z=0$. The action is branching at turning
points, $S(z)-S(z_t)\propto (z-z_t)^{3/2}$. Its behavior near
caustics in a multidimensional system is very similar (see below),
with $z-z_t$ corresponding to the distance from the caustic.

Since neighboring Hamiltonian trajectories (\ref{eom}),
(\ref{initial_general}) touch each other on a caustic, the one-to-one
correspondence between the coordinates $x_1$, $x_2$, $x_3$ on the
trajectory and the parameters $t$, $x_1(0)$, $x_2(0)$
breaks. Therefore the equation for a caustic has the form
\begin{equation}
\label{caustic}
J({\bf r})=0,\; J({\bf r})={\partial( x_1,x_2,x_3)\over \partial
(x_1(0),x_2(0),t)}.
\end{equation}
The Jacobian $J({\bf r})$ can be related in a standard way to the
prefactor $D({\bf r})$, which in turn is determined by the first-order
(in $\hbar$) correction to the action $-iS^{(1)}$, $D({\bf r})=\exp[
S^{(1)}({\bf r})]$. The equation for $S^{(1)}({\bf r})$ can be
obtained from the Schr\"odinger equation by seeking the wave function
in the form $\psi = \exp(iS)$, with $S=S^{(0)}-i S^{(1)}$. This gives
$2{\bf v}\nabla S^{(1)}= -{\bbox\nabla} {\bf v}$, where $m{\bf v} =
{\bbox\nabla} S^{(0)} + (e/c){\bf A}$.  The vector ${\bf v}$ gives the
velocity on the Hamiltonian trajectory (\ref{eom}). Taking into
account that ${\bf v}\nabla S^{(1)}({\bf r})
\equiv dS^{(1)}/dt$ and that ${\bbox\nabla} {\bf v} = d\ln J({\bf
r})/dt$, where the time derivatives are taken along the trajectory, we
obtain
\begin{equation}
\label{prefactor}
D({\bf r}) = {\rm const}\times [J({\bf r})]^{-1/2}.
\end{equation}
It follows from Eqs.~(\ref{caustic}), (\ref{prefactor}) that the
prefactor $D$ diverges on caustics, and the WKB approximation does not
apply there\cite{Berry}. 

There are both formal and physical distinctions between caustics for
tunneling trajectories with and without a magnetic field. For $B\neq
0$, the trajectories are complex, and Eq.~(\ref{caustic}) specifies a
complex surface. This surface intersects the real 3D space along a
line. In contrast, for $B=0$, because of time-reversal symmetry,
tunneling trajectories lie in real configuration space. In this
symmetric case Eq.~(\ref{caustic}) specifies a surface
\cite{Schmid} rather than a line in the 3D space. As we show
below, this distinction leads to observable consequences.

\subsection{Local analysis near caustics}

The analysis of the wave function and branching of the action $S({\bf
r})$ at complex caustics in the magnetic field is similar to that for
caustics in real space, including turning points in the 1D case
\cite{Berry}. Near a caustic, it is convenient to change to the
variables $x',~y'$, and $z'$ which are locally parallel and
perpendicular to the caustic surface, respectively (we set $z'=0$ on
the caustic). Since a caustic is an envelope of the Hamiltonian
trajectories (\ref{eom}), the normal to the caustic component of the
velocity is $v_{z'}=0$ for $z'=0$. However, for $B\neq 0$ the normal
component of the momentum is not equal to zero. Therefore the wave
function near a point ${\bf r}_{\rm caust}$ on the caustic can be
sought in the form
\begin{equation}
\label{loc_var}
\psi({\bf r}_{\rm caust}+{\bf r}')=e^{i{\bf p}_{\rm caust}{\bf r}'}
\phi(z';{\bf r}_{\rm caust}),
\end{equation}
where ${\bf p}_{\rm caust}$ is the momentum at the point ${\bf r}_{\rm
caust}$ (note that ${\bf r}_{\rm caust}$ is a 2D complex vector,
$z'_{\rm caust}\equiv 0$). We assume that the dependence of ${\bf
p}_{\rm caust}$ on ${\bf r}_{\rm caust}$ (i.e., along the caustic) is
smooth, and the dependence of $\phi$ on ${\bf r}_{\rm caust}$ is much
more smooth than on $z'$.  Generally, ${\bf p}_{\rm caust}$ is complex
even where the caustic goes through real space, Im~${\bf r}_{\rm
caust}=0$. Therefore the classical trajectory of the escaped particle
does not go through the caustic, in contrast to the case of zero
magnetic field, cf. Fig.~\ref{fig:caus}.

Because $v_{z'}=0$, the equation for $\phi(z';{\bf r}_{\rm caust})$, which
follows from the 3D Schr\"odinger equation with a magnetic field,
coincides with the 1D Schr\"odinger equation near a turning point
\begin{equation}
\label{1d_schr}
\left[-
{1 \over 2m}{d^2 \over d{z'}^2}+ U'_{z'}({\bf
r}_{\rm caust})z'\right]\phi(z';{\bf r}_{\rm caust})=0
\end{equation}
[here, $U'_{z'} \equiv \partial U/\partial z'$]. The boundary
conditions to this equation are discussed below. 

The function $\phi$ is singled-valued. It is given by a linear
combination of the Airy functions
\cite{LandauQM}. For comparatively large $|z'|$ (but still close to
the caustic) it becomes a linear combination of the functions
\begin{equation}
\label{airy}
w_{1,2}=(z')^{-1/4}\exp\left[\mp i\alpha {z'\,}^{3/2}\right],
\end{equation} 
with $\alpha \equiv \alpha({\bf r}_{\rm caust})= (2/3)[-2mU'_{z'}]^{1/2}$. To
make the functions $w_{1,2}$ uniquely defined, we have to make a cut
on the complex $z'$-plane. Our choice of the cut is shown in
Fig.~\ref{fig:stokes}.

With account taken of Eq.~(\ref{loc_var}), we find that the action
near the caustic is
\begin{eqnarray}
\label{increment}
S(x',y',z')\approx S(x',y',0) + (p_{\rm caust})_{z'}z' + \alpha {z'\,}^{3/2},
\end{eqnarray}
with an appropriately chosen branch of ${z'\,}^{3/2}$.

Another way to understand Eq.~(\ref{increment}) is based on the
analysis of the set of the Hamiltonian trajectories (\ref{eom}).
Because the caustic is an envelope of the trajectories and $v_{z'}=0$
on the caustic, $z'$ is quadratic in the increments $\delta
x_{1,2}(0),\delta t$ of the parameters of the set.  Therefore $\delta
x_{1,2}(0), \delta t$ are nonanalytic in $z'$, as is also the action
$S$.  Taking into account cubic terms in $\delta x_{1,2}(0),\delta t$
we obtain (\ref{increment}). The coefficients in $S$ can be expressed
in terms of the derivatives of $S,{\bf r}$ calculated along the
trajectories over $x_{1,2}(0),t$.

\subsection{Choosing the action branches}

Eq.~(\ref{1d_schr}) describes how the WKB solutions, which correspond
to different branches of the action $S$, connect on the caustic.  Of
interest for the problem of tunneling escape is the caustic where
there are connected the tails of the intrawell wave function and the
outgoing wave packet for the escaped particle. From (\ref{caustic}),
on this caustic there is a point through which there goes the
Hamiltonian trajectory (\ref{eom}), (\ref{initial_general}) for
escape, with the initial coordinates $x_{1,2}(0)$ given by the
condition (\ref{exit}) of arriving, ultimately, at the classical
outgoing trajectory ${\bf r}_{\rm cl}$, cf. Fig.~\ref{fig:caus}.

\begin{figure}
\begin{center}
\epsfxsize=3.0in                
\leavevmode\epsfbox{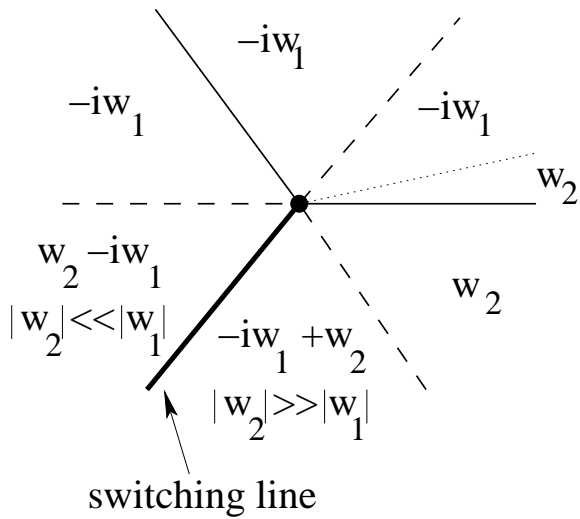}
\end{center}
\caption{The function $\phi(z')$ (\protect\ref{1d_schr}) for large $|z'|$ on 
the complex $z'$-plane perpendicular to the caustic surface.  The
dashed lines, $\arg z' = (2n+1)\pi/3$, show the Stokes lines where the
ratio of the functions $|w_2/w_1|$ (\protect\ref{airy}) reaches its
maximum or minimum. The anti-Stokes lines (solid lines), $\arg
z'=2n\pi/3$, are the lines where $|w_1|=|w_2|$. At these lines the
ratio $|w_2/w_1|$ changes from exponentially large to exponentially
small with varying $\arg z'$. The coefficients are found from the radiation
boundary condition following the Stokes prescription
\protect\cite{Berry}. The dotted line shows the branch cut.}
\label{fig:stokes}
\end{figure}

In general, a caustic can be thought of as a ``mirror'', which partly
reflects the wave packet. The boundary condition to
Eq.~(\ref{1d_schr}) for the tunneling escape problem is the
``radiation condition''. Very far from the potential well the solution
of the full Schr\"odinger equation is a semiclassical wave packet
moving in real space away from the well. Respectively, the wave
function far from the well has a form $\psi({\bf r}) \propto
\exp[iS({\bf r})]$. This solution has to be continued to the
caustic, which means that there is a range of directions in the
complex $z'$-plane not too close to the caustic ($|\alpha {z'\,}^{3/2}|
\gg \hbar$) where the wave function is described by only one exponential
$\exp[iS({\bf r})]$. Away from this range, the wave function is a
combination of two waves. This is again similar to a 1D problem, where
behind the turning point $z_t$ the wave function for real $z$ is a
propagating wave, whereas before $z_t$ there is a wave with an
exponentially decaying amplitude coming from the intrawell state and
the wave reflected back to this state. 

To connect the tails of the
wave functions near the caustic in our problem one can use the
explicit solution of Eq.~(\ref{1d_schr}) with the radiation boundary
condition, as in the 1D case \cite{LandauQM}.

An alternative way to see how the boundary condition works, which also
allows us to reveal the specific feature of tunneling in a magnetic
field, is to follow the transformation of the wave function $\phi$ for
$|\alpha {z'\,}^{3/2}| \gg \hbar$ as the argument of $z'$ varies by
$2\pi$. This analysis is based on the notion of the Stokes and
anti-Stokes lines \cite{Berry,Stokes,Berry-Mount}. We count $\arg z'$ off
from $\arg \alpha^{2/3}$. Then, for the choice of the cut in
Fig.~\ref{fig:stokes} and with the functions $w_{1,2}$ given by
Eq.~(\ref{airy}), the Stokes lines are the rays $\arg z' =
(2n+1)\pi/3$ with $n=0,1,2$. On the Stokes lines ${\rm
Re}\,{z'\,}^{3/2} = 0$ and the ratio $|w_2/w_1|$ is extremal (maximal
or minimal). The anti-Stokes lines are the rays $\arg z' = 2n\pi/3$
with $n=0,1,2$, where $|w_2/w_1|=1$.

From the radiation boundary condition, there is a range $\Delta$ of
$\arg z'$ where $\phi(z')$ is given by only one of the functions
$w_{1,2}(z')$,  not by a superposition of $w_1$ and
$w_2$. This condition is physically meaningful provided the
corresponding $w_i$ is exponentially {\it small} compared to $w_{3-i}$
in a part of the range $\Delta$. [Otherwise the condition $\phi(z') =
{\rm const}\times w_i$ is not a limitation, in the WKB approximation.]
It follows from the analysis below that $\Delta$ includes one of
the rays $\arg z'=0$ or $\pi$, along which $\phi(z')$ is oscillating
as $\exp(i \alpha {z'\,}^{3/2})$ or $\exp(-i\alpha {z'\,}^{3/2})$.

For concreteness, we will assume that $\Delta$ contains the
semi-axis $\arg z'=0$, and that between $\arg z'=0$ and the cut in
Fig.~\ref{fig:stokes} $\phi(z') = C\, w_2(z') \propto
\exp[i|\alpha|{z'\,}^{3/2}]$, where $C$ is a constant (it is not
incorporated into Fig.~\ref{fig:stokes}). 


Since the function $\phi(z')$ is single-valued, if we cross the cut in
Fig.~\ref{fig:stokes} by incrementing $\arg z'$,
$\phi(z')$ becomes equal to $-iCw_1(z')$. It remains exponentially
small as $\arg z'$ grows up to $2\pi/3$, including the Stokes line
$\arg z'=\pi/3$. Then behind the anti-Stokes line at $\arg z' =
2\pi/3$, the function $w_1$ becomes exponentially big. It is important
that, on the Stokes line $\arg z' = \pi$, one has to take into account
the admixture to $\phi$ of an exponentially small term $\propto
w_2(z')$. This can be seen from the explicit solution for $\phi$. The
need to incorporate this term can also be understood by noticing that,
when we increment $\arg z'$ by $2\pi$, we have to recover the original
asymptotic form of $\phi$. This latter argument explains the
coefficient at $w_2$ in Fig.~\ref{fig:stokes}.

On the anti-Stokes line $\arg z' = 4\pi/3$, the values of $|w_1|$ and
$|w_2|$ become equal to each other, and $\phi(z')$ is primarily
determined by $w_2$ for larger $\arg z'$. After $\arg z'$ crosses the
Stokes line $5\pi/3$, the exponentially small term $w_1$ in $\phi(z')$
disappears, according to the Stokes prescription \cite{Berry}. 

The wave function which connects to the outgoing wave on the caustic
is that of the metastable state. Since the asymptotic behavior of
$\phi(z')$ for large $|z'|$ is given by $w_2(z')$, the asymptotic
behavior of the wave function of the metastable state within the range
$2\pi/3 < \arg z' < 4\pi/3$ is given by $w_1(z')$. We note that, for
the radiation boundary condition, the switching between the wave
functions occurs only on one of the anti-Stokes lines.

\subsection{Switching between the branches of the wave function}

Switching between the WKB wave functions of the localized state and
the outgoing wave packet is an important observable consequence of the
analysis in the previous subsection. It is due to branching of the WKB
action. The switching manifold starts on the caustic and is given by
the condition Im~$S_1({\bf r}) = {\rm Im} S_2({\bf r})$, where
$S_{1,2}$ are the actions for the corresponding WKB branches. On the
opposite sides of the switching manifold one of the WKB wave functions is
exponentially bigger than the other.

In the presence of a magnetic field, caustics go through real space
along the lines given by the condition $J({\bf r}) = 0$, Im~${\bf r}
=0$, with ${\bf r}\equiv {\bf r}(x_{1,2}(0),t)$ being a point on the
Hamiltonian trajectory (\ref{eom}), (\ref{initial_general}). The
switching manifold in real space is a surface which starts from the
caustic line and goes away from it in one direction. Although the wave
function is continuous on this surface, the derivative of its
logarithm sharply changes from ${\bbox\nabla}S_1$ to
${\bbox\nabla}S_2$.

The exit point ${\bf r}$ in the configuration space where the escaped
particle emerges from under the barrier is determined by the
intersection of the classical escape trajectory ${\bf r}_{\rm cl}(t)$
and the switching surface. This point can be found from the global
analysis of the WKB wave function. It does not lie on the caustic, nor
is it given by the condition ${\bf p}=0$ or equivalently, $U({\bf
r})=E$. In a 2D system the caustic pierces real space at a point, and
the switching surface becomes a line. An example of the caustic, the
switching line, and the exit point for a 2D system is shown in
Fig.~\ref{fig:time}.

In the absence of a magnetic field, as we mentioned before, the
caustic for the tunneling trajectories is a surface rather than a line
in real configuration space. The switching surface in real space
should then coincide with the caustic. The exit point is the turning
point ${\bf p = 0}$. The whole topology thus differs qualitatively
from that for $B\neq 0$.

\section{Tunneling from a correlated 2D electron system}  
\label{sec:correlated}

We now apply the general approach to tunneling from a correlated 2D
electron system transverse to a magnetic field and illustrate the
occurrence of the singularities discussed above. We will use the
simple but nontrivial model of the electron system discussed in
Sec.~\ref{sec:model} and described by Eqs.~(\ref{U(z)}),
(\ref{U_0(z)}), and its generalization to the case where the in-plane
symmetry is broken. We will assume that the magnetic field is parallel
to the electron layer, and choose the $y$-axes along the field ${\bf
B}$.

\subsection{A model with in-plane symmetry}
\label{subsec:symmetry}

For an electron in the potential (\ref{U(z)}),
(\ref{U_0(z)}), classical motion along the ${\bf B} \parallel
\hat{\bf y}$ axis is decoupled from the motion in the $(x,z)$-plane. 
The WKB tunneling problem then becomes two-dimensional, with complex
classical trajectories (\ref{eom}) lying in this plane.  The
Hamiltonian equations are linear, and we can find the trajectories
explicitly. We can also explicitly find the tunneling exponent and
analyze \cite{Barabash-2000} its dependence on the two dimensionless
parameters $\omega_0\tau_0$ and $\omega_c\tau_0$, where $\tau_0$ is
the tunneling time in the absence of the magnetic field,
$\tau_0=2mL/\gamma$. The expression for the tunneling exponent also
follows as a limiting case from the result of the next
subsection. Here we will discuss the structure of the WKB action.

\begin{figure}
\begin{center}
\epsfxsize=3.0in                
\leavevmode\epsfbox{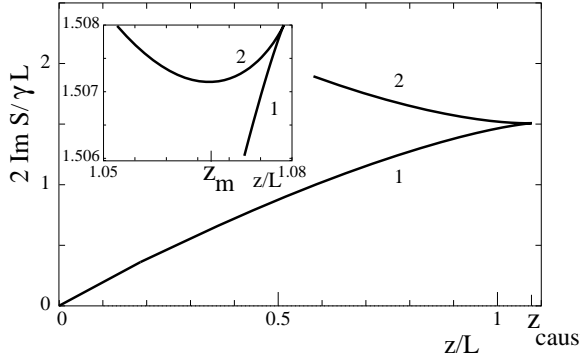}
\end{center}
\caption{Two branches of the action  on 
the symmetry axis $x=0$ as a function of the tunneling coordinate $z$
before the branching point, for the symmetric model (\ref{U(z)}) and
same parameter values as in Fig.~\ref{fig:time}.  The vicinity of the
cusp $z_{\rm caust}$ is zoomed in the inset to show that the upper
branch is nonmonotonic. Its extremum at $z_m$ lies on the classical
trajectory of the escaping particle shown in
Fig.~\ref{fig:time}(b). However, the particle emerges from the barrier
for $z>z_m$ and $x\neq 0$.}
\label{fig:beak}
\end{figure}

The symmetry of the potential $U(x,y,z)=U(\pm x,\pm y,z)$ in
(\ref{U(z)}) gives rise to a specific symmetry of the set of the
Hamiltonian trajectories,
\begin{equation}
\label{symmetry}
t\to -t^{*},\; x\to -x^{*},\; z\to z^{*},\; S\to -S^{*},
\end{equation}
and of the singularities of this set. In particular, the caustic where
there are connected the outgoing wave packet and the intrawell wave function
goes through real plane $(x,z)$ at a point $x_{\rm caust}=0$ on the
symmetry axis. The $z$-coordinate of this point depends on the
magnetic field, with $z_{\rm caust}=L$ for $B=0$. The form of the
action for $z \leq z_{\rm caust}$ near the caustic in the symmetry
plane $x=0$ is shown in Fig.~\ref{fig:beak}. As seen from the inset,
the slope of the action is $\partial {\rm Im}~S/\partial z >0$ at
$z_{\rm caust}$, in contrast to the 1D case where the slope is equal
to zero at the turning point $z_t$. We note that the branches 1 and 2
are formed by the trajectories that go through real space at times
-Im~$t$ being, respectively, smaller and larger than -Im~$t_{\rm
caust}$ for the same trajectories [$t_{\rm caust}$ is given by
Eq.~(\ref{caustic})].

Using the explicit form of the trajectories (\ref{eom}) with the
initial conditions (\ref{initial}), one can find the complex caustic
$z'(x,z)=0$ near $x=0,z=z_{\rm caust}$. It has the form $z-z_{\rm
caust}= iC'x$ with real $C'/B> 0$, cf. Fig.~\ref{fig:caus}. It is seen
from Fig.~\ref{fig:beak} that the singular parts of Im~$S_{1,2}$ behave as
$\mp (z_{\rm caust}-z)^{3/2}$ near the caustic. Therefore we can
choose the coordinate $z'$ that gives the distance from the caustic as
$z-z_{\rm caust}-iC'x$. The branching behavior near the caustic is
then described by Fig.~\ref{fig:stokes}. The range $\pi\leq
\arg z'<2\pi$ corresponds to real $z$ and real positive $x$.

Close to the caustic, only one branch of the action
describes the wave function for negative $x$ and real $z$ (the upper
half of the complex-$z'$ plane). For positive $x$, we should keep both
branches, and depending on $x,z$, the WKB wave function is given by
the branch with the smaller Im~$S$. Near the switching line where
Im~$S_1= {\rm Im}~S_2$ the total wave function is given by a linear
combination of the two WKB solutions. 

In Fig.~\ref{fig:beak} for $z\leq z_{\rm caust}$,  the action branch 1 
describes the tail of the intrawell wave function, and the branch 2
corresponds to the wave ``reflected'' from the caustic.  The branch 2
is nonmonotonic in $z$ for $x=0$, with a minimum at $z_m<z_{\rm
caust}$. By symmetry (\ref{symmetry}), the momentum component $p_x$ is
real for $x=0$, whereas $p_z=0$ at $z_m$.  Therefore the point
$z=z_m$, $x=0$ belongs to the classical escape trajectory shown in
Fig.~\ref{fig:time}(b). Again by symmetry, this is also the point
where the escape trajectory comes closest to the well at
$z=0$. However, this is not the exit point for the
tunneling particle in the configuration space. Indeed, the wave
function at this point is determined by the branch 1, because Im~$S_1
< {\rm Im}~S_2$.

\begin{figure}
\begin{center}
\epsfxsize=3.3in
\leavevmode\epsfbox{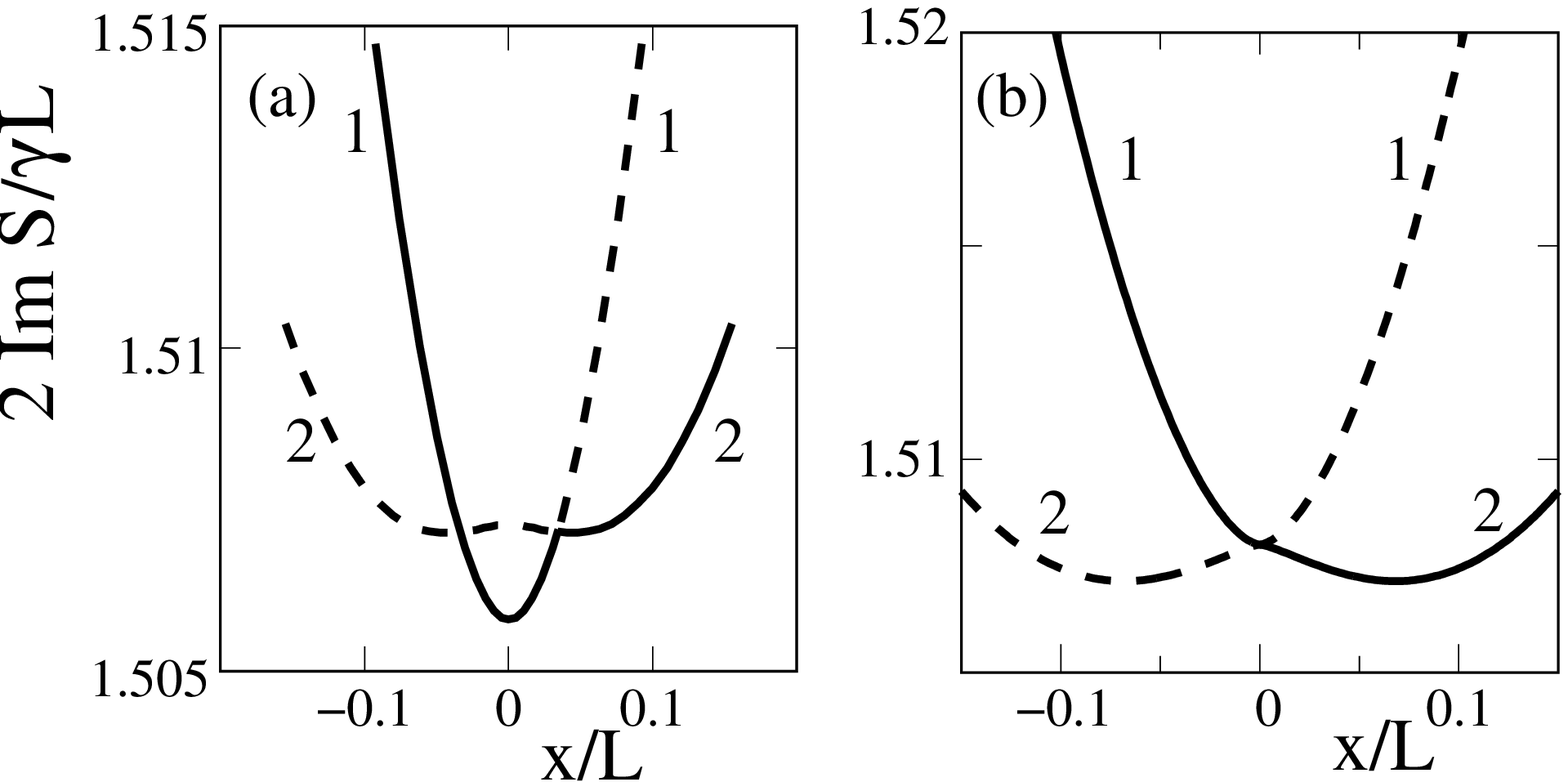}
\end{center}
\begin{center}
\epsfxsize=3.0in                
\leavevmode\epsfbox{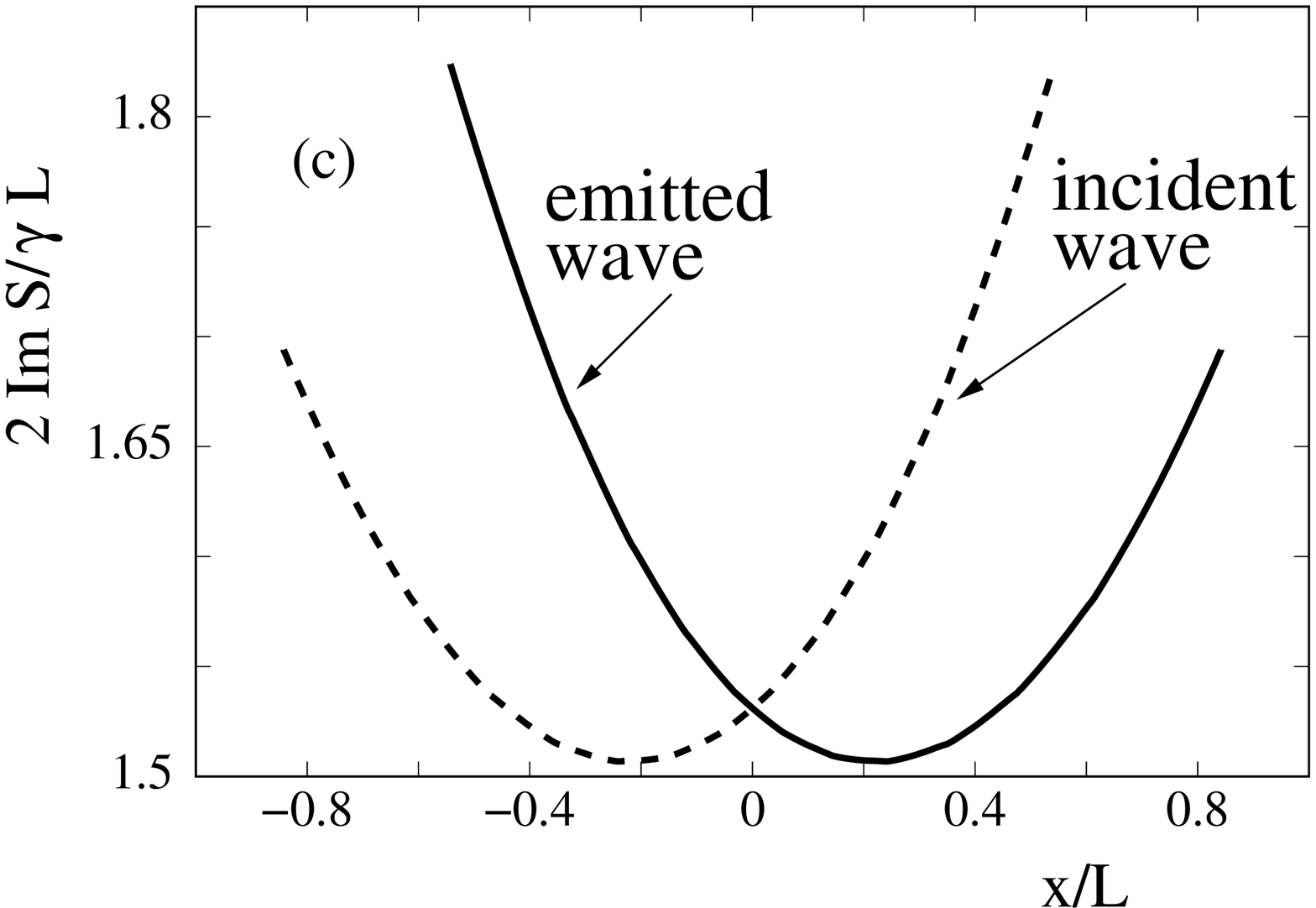}
\end{center}
\caption{Cross-sections of the function ${\rm Im}~S$ by the plane $z=$~const 
in the case of tunneling in a
symmetric potential (\ref{U(z)}), for (a) $z_m<z<z_{\rm caust}$,
(b) $z=z_{\rm caust}$, and (c) $z>z_{\rm caust}$.  The parameter values
are the same as in Figs.~\ref{fig:time},
\ref{fig:beak}. The solid lines show the branches of Im~$S$ that
determine the exponent of the WKB wave function. The minima of the
branch 2 lie on the classical trajectory shown in
Fig.~\ref{fig:time}(b).}
\label{fig:sections}
\end{figure}

Cross-sections of the action surfaces by planes $z=$~const are shown
in Fig.~\ref{fig:sections}.  For $z\leq z_{\rm caust}$, both branches
Im~$S_{1,2}$ are symmetrical in $x$. However, the branch 2 is
nonmonotonic in $x$ for $z>z_m$. It has a local {\it maximum} at $x=0$
and two symmetrical {\it minima}. These minima lie on the classical
trajectory shown in Fig.~\ref{fig:time}(b). For $z=z_m$, the maximum
and the minima merge together. We note that at this point
Im~$S_2\propto x^4$ near the minimum.

Behind the caustic in real space, $z>z_{\rm caust}$, the action
$S(x,z)$ on one of the two branches is equal to $-S^*(-x,z)$ on the
other branch (cf. Fig.~\ref{fig:sections}). At their minima with
respect to $x$, the values of Im~$S(x,z)$ are independent of
$z$. These minima lie on the escape classical trajectory, as seen from
the comparison with Im~$S$ for $z< z_{\rm caust}$. The branches of the
action describe the wave packets incident on the barrier from large
$z$ and the emitted wave packet. Only the latter is physically
meaningful for the problem of tunneling escape.

As discussed above, switching between the action branches occurs for positive
$x$ where the branches of Im~$S$ cross each other. The resulting
action, which determines the WKB wave function, is shown in
Fig.~\ref{fig:sections} by solid lines. The switching line thus
obtained coincides near the caustic with the anti-Stokes switching line
discussed in the previous section.

The escaped particle can be observed as a semiclassical object if
${\rm Im}~S_2({\bf r}_{\rm cl})<{\rm Im}~S_1({\bf r}_{\rm cl})$. It
``shows up'' at the point where the classical escape trajectory
intersects the switching line, cf. Fig.~\ref{fig:time}.  The exit
point is located for $x> 0$ even though the potential (\ref{U(z)}) is
symmetric. This is a consequence of the symmetry-breaking by a
magnetic field.

\subsection{A  non-symmetric model}
\label{sec:asymmetry}

The problem considered in the previous section for the quadratic in
$x,y$ potential $U({\bf r})$ (\ref{U(z)}) can be solved
differently. The trick is \cite{Caldeira} to make a canonical
transformation to the new coordinate $p_x$ and the conjugate
momentum $-x$. The kinetic energy then becomes
$[m^2\omega_0^2x^2 + p_z^2]/2m$ and is independent of the new
coordinates and the magnetic field. The time-reversal symmetry is thus
``restored'', and the problem is mapped onto the standard problem of
tunneling in the 2D potential $U_0(z) + (p_x+m\omega_cz)^2/2m$.

The general method discussed in this paper is not limited to
potentials with these special properties. In this subsection we
illustrate how this method works where variables do not separate. To
this end, we consider a 2D problem of tunneling transverse to the
field ${\bf B\parallel \hat y}$ in the potential
\begin{equation}
\label{U(x,z)}
U(x,z)= {1\over 2}m\omega_0^2x^2 + \mu xz+{\gamma^2\over
2m}\left(1-{z\over L}\right) \; (z>0),
\end{equation}
which differs from the potential discussed earlier by the term $\mu x
z$. In the problem of tunneling from a 2D electron system, this term
mimics the dependence of the tunneling potential on the displacements
of neighboring electrons, see Appendix~\ref{app1}.

\begin{figure}
\begin{center}
\epsfxsize=3.3in                
\leavevmode\epsfbox{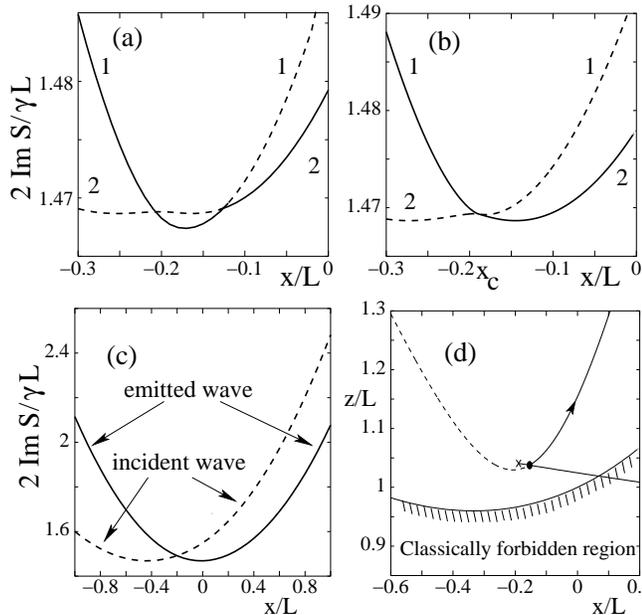}
\end{center}
\caption{Cross-sections of Im~$S(x,z)$  by the plane $z=$~const 
in the case of tunneling in the asymmetric 2D potential $U(x,z)$
(\ref{U(x,z)}), for (a) $z_m<z<z_{\rm caust}$, (b) $z=z_{\rm caust}$,
and (c) $z>z_{\rm caust}$.  The values of $\omega_c\tau_0$ and
$\omega_0\tau_0$ are the same as in Figs.~\ref{fig:time},
\ref{fig:beak}, and the dimensionless asymmetry parameter 
$4\mu mL^2/\gamma^2 =1/2$. The solid lines show the branches of Im~$S$
that determine the exponent of the WKB wave function. The minima of
the branch 2 lie on the classical escape trajectory shown in (d).  The
cross in (d) marks the branching point of the function ${\rm Im}~S$
where the caustic goes through real space. The switching line (thin
solid line) starts at the branching point.}
\label{fig:asym_action}
\end{figure}

\begin{figure}
\begin{center}
\epsfxsize=3.3in                
\leavevmode\epsfbox{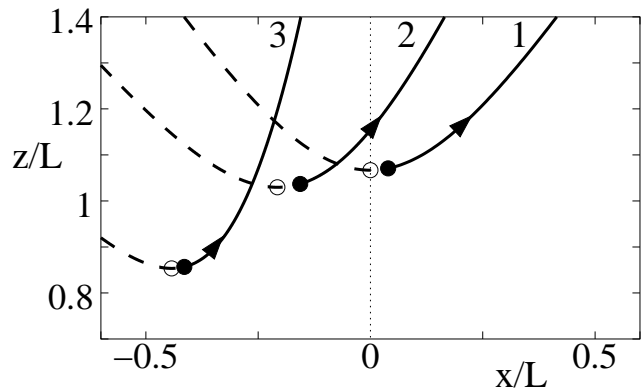}
\end{center}
\caption{Escape trajectories for the dimensionless asymmetry parameter 
$4\mu mL^2/\gamma^2=0,0.5,1.5$. The filled circles are exit points, and
the open circles are the points where the trajectory is most close to the
localized state.}
\label{fig:asym_tr}
\end{figure}

The term $\mu x z$ breaks the symmetry of the Hamiltonian trajectories
(\ref{symmetry}). However, the Hamiltonian equations (\ref{eom}) are
still linear, and we explicitly solved them.  The results for Im~$S$
and the classical escape trajectory obtained using the initial
conditions (\ref{initial}) are shown in Figs.~\ref{fig:asym_action},
\ref{fig:asym_tr}.  Because of broken symmetry, the branching point of
the action in real space, where the caustic of the set of trajectories
(\ref{eom}) goes through real plane $(x,z)$, lies at $x_{\rm
caust}\neq 0$. It is marked by the cross on
Fig.~\ref{fig:asym_action}(d). Its position depends on $\mu$ and other
parameters of the system. Similarly, the time where the caustic
crosses the real space has both real and imaginary part, in contrast
to the case $\mu=0$ where it was purely imaginary, see
Fig.~\ref{fig:time}.

For $\mu\neq 0$, the surfaces Im~$S(x,z)$
become asymmetric. The general structure of the solution, however,
remains the same as in the case $\mu=0$. This can be seen by comparing
the cross-sections of the action in Figs.~\ref{fig:sections} and
\ref{fig:asym_action}. In both figures, the  cross-sections in 
(a), (b), and (c) refer to the planes $z<z_{\rm caust}, \, z= z_{\rm
caust}$, and $z>z_{\rm caust}$, respectively. As in the symmetrical
case, the branches 1 and 2 for $\mu\neq 0$ are formed by the
trajectories in complex time $t$, with -Im~$t$ being, respectively,
smaller and bigger than -Im~$t_{\rm caust}$. At the point $x_{\rm
caust},z_{\rm caust}$, the branches 1 and 2 touch each other.

For $z<z_{\rm caust}$ there occurs switching between the branches. It
can be analyzed in the same way as for $\mu=0$. The WKB wave function
is determined by the branch in Fig.~\ref{fig:asym_action} shown with
the solid line. The switching line starts from the branching point
$x_{\rm caust},z_{\rm caust}$ and goes in the direction of positive $x$.

The branch 2 has two minima as a function of $x$ for $z_m<z< z_{\rm
caust}$, which are asymmetric for $\mu \neq 0$. However, their depths,
i.e. the minimal values of Im~$S$ on this branch, remain equal to each
other and are the same in all cross-sections $z=$~const. These minima
lie on the classical trajectory along which the electron escapes. At
$z=z_m$ they merge together, and Im~$S$ becomes quartic in $x$ near
the minimum.  The value $z_m$ shows how close the escape trajectory
comes to the localized intrawell state.

The classical trajectory becomes observable in configuration space
once it crosses the switching line. The shape of the trajectory and
the exit point for several values of the asymmetry parameter $\mu$ are
shown in Fig.~\ref{fig:asym_tr}. The outgoing wave packet is Gaussian
near the maximum (Im~$S$ is parabolic near the corresponding
minimum). 

By solving the Hamiltonian equations (\ref{eom}) for the potential
(\ref{U(x,z)}), we obtained the following expression for the
tunneling exponent (\ref{answer}),
\begin{equation}
R=2\gamma L [\tau_{\rm rd}+\nu_0\kappa(\tau_{\rm rd})]
\label{R_asymm}
\end{equation}
Here, $\tau_{\rm rd}$ is the imaginary part of the time to reach the
classical escape trajectory [see Fig.~\ref{fig:time}(a)] in the units
of that same time for $B=\mu=0$, which is given by
$\tau_0=2mL/\gamma$. Along with the function $\kappa$ in
(\ref{R_asymm}), $\tau_{\rm rd}$ can be found from the equation

\wid
\begin{eqnarray}
\label{kappa}
\kappa(\tau_{\rm rd})\equiv&& {\nu_0^2 (\tau_{\rm rd} \cos \nu\!_{_-} 
\tau_{\rm rd} - {\nu\!_{_-}}\!\!^{-1} \sin \nu\!_{_-} \tau_{\rm rd}) + 
{\nu\!_{_-}}\!^2 \cos \nu\!_{_-} \tau_{\rm rd} - \nu\!_{_-} (1- \nu_0+
 \nu_0\tau_{\rm rd})\sin\nu\!_{_-}\tau_{\rm rd} \over
 \tilde{\mu}^2(\nu_0\cos\nu\!_{_-}\tau_{\rm
 rd}-\nu\!_{_-}\sin\nu\!_{_-}\tau_{\rm rd})}\nonumber\\
 &&={(\nu_0^2\tau_{\rm rd} -{\nu\!_{+}}\!^2)\cosh \nu\!_{+}\tau_{\rm
 rd}+\nu\!_{+}(1-\nu_0+\nu_0\tau_{\rm
 rd}-\nu_0^2/{\nu\!_{+}}\!^2)\sinh\nu\!_{+}\tau_{\rm rd} \over
 \tilde{\mu}^2 (\nu_0\cosh\nu\!_{+}\tau_{\rm
 rd}+\nu\!_{+}\sinh\nu\!_{+}\tau_{\rm rd})}.
\end{eqnarray}
\nar
Here, $\tilde\mu=\mu\tau_0^2/m$ is the dimensionless asymmetry
parameter. The dimensionless frequencies $\nu_0=\omega_0\tau_0$,
$\nu_c= \omega_c\tau_0$, and $\nu = (\nu_c^2 + \nu_0^2)^{1/2}$
characterize the motion under the barrier,
${\nu\!_{_\pm}}\!^2=\pm\nu^2/2+\sqrt{\nu^4/4+\tilde\mu^2}$.

\begin{figure}
\begin{center}
\epsfxsize=3.3in                
\leavevmode\epsfbox{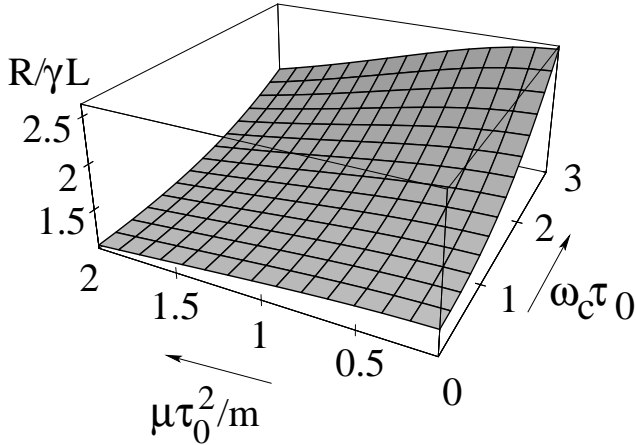}
\end{center}
\caption{The tunneling exponent $R$ as a function of the magnetic field and
the asymmetry parameter $\mu$ in the model (\protect\ref{U(x,z)}), for
$\omega_0\tau_0=1.2$. The function $R$ is even in $\mu$.}
\label{fig:3DR}
\end{figure}

The tunneling exponent $R$ depends on the interrelation between the
in-plane electron dynamics, which is characterized by the frequency
$\omega_0$, the cyclotron frequency $\omega_c$, the tunneling time
$\tau_0$, and the asymmetry-induced mixing of in-plane and
out-of-plane motions $\tilde\mu$.  The dependence of $R$ on the
asymmetry parameter and on the magnetic field is shown in
Fig.~\ref{fig:3DR}. For $\omega_c=\tilde\mu = 0$, we have $\tau_{\rm
rd}=1$, and $R=4\gamma L/3$. The exponent $R$ increases with the
magnetic field. A qualitative result of the in-plane
confinement\cite{Barabash-2000} is that it {\it eliminates} the
localization in the linear potential $U_0(z)$ due to the magnetic
barrier\cite{Andreimagn,Popov}, i.e. the divergence of $R$ for
$\omega_c\tau_0 >1$.  The effect of the magnetic field on $R$ becomes
small for strong confinement, $\omega_0\tau_0 \gg 1$ and $\omega_0\gg
\omega_c$.  

The asymmetry results in lowering of the tunneling barrier for $B=0$
and the corresponding increase of the tunneling rate. This can be
qualitatively understood, since a displacement in the $x$-direction
with $\mu x < 0$ increases the effective force in the $z$-direction
which pulls the electron away from the layer. For small asymmetry,
$\tilde\mu \ll 1$, the correction to $R$ is quadratic in
$\tilde\mu$. In the limit of a thin and high tunneling barrier for
$x=0$ or soft in-plane vibrations, where $\omega_0\tau_0 \ll 1$,
tunneling is most likely to occur in the barrier, which is
adiabatically prepared by the in-plane displacement $x$ (the
``completely adjusted'' barrier\cite{AP}). The expression for $R$
takes a form
\begin{equation}
\label{simpleR}
R=2\gamma
L(3\nu_0/\tilde\mu^2)^{1/3} \quad (\omega_0\tau_0\ll 1),
\end{equation}
it depends on the in-plane frequency $\omega_0$ and on $\mu$
nonanalytically. The role of the asymmetry increases with the magnetic
field, as seen from Fig.~\ref{fig:3DR}.

In terms of comparison with the currently available experimental data
on tunneling from a correlated many-electron system \cite{Andreimagn},
of utmost interest is the situation where the asymmetry is small. The
observed dependence of $R$ on $B$ did not show the divergence expected
for an unconfined electron.  The simple model (\ref{U(z)}) provides a
qualitative explanation of the experiment \cite{Barabash-2000}.
An excellent quantitative agreement, without adjustable parameters,
was achieved by incorporating the curvature of the potential $U_0(z)$
due to electron correlations\cite{Dykman-01}.

\section{The path-integral formulation in a magnetic field}
\label{sec:instanton}
In the absence of a magnetic field, the problem of tunneling decay is
often considered using the instanton technique
\cite{Langer,Coleman}. This technique applies if the potential 
well is parabolic near the minimum and thermalization inside the well
occurs much faster than escape from the well (in the case of 2D
electron systems, both conditions are often violated
\cite{Dykman-01}). Because
the Schr\"odinger equation for metastable states has to be solved with
the radiation boundary condition, the energies of these state acquire
small imaginary parts, and so does the partition function $Z$. The
escape rate $W$ for finite temperatures is simply related to Im~$Z$,
\begin{equation}
\label{rate}
W\approx 2 T\, {\rm Im}~Z /{\rm Re}~Z
\end{equation}
(we have set $k_B=1$).

The partition function is given by the integral over periodic paths
${\bf r}(\tau)$ in imaginary time \cite{Feynman},
\begin{equation}
\label{stsum}
Z=
\int_{{\bf r}(0)={\bf r}(\beta)} {\cal D} {\bf r}(\tau)
\exp\left[-S_E[{\bf r}(\tau)]\right], 
\end{equation}
where $\beta=T^{-1}$ and $S_E$ is the Euclidean action (the action in
imaginary time). It is real for $B=0$ and  for real trajectories ${\bf
r}(\tau)$.

The general expressions (\ref{rate}), (\ref{stsum}) should also apply
in the presence of a magnetic field. However, the Euclidean action for
an electron
\begin{equation}
\label{euclid_action}
S_E=\int_0^{\beta} d\tau \left[ {m\over 2}
\left(d {\bf r} \over d\tau \right)^2 +U({\bf r}) +i {e\over c} {\bf A} ({\bf
r}) {d{\bf r}\over d\tau} \right]
\end{equation}
is now complex. Therefore the standard way \cite{Langer,Coleman} of
evaluating the escape rate has to be revised, except for special
symmetric cases like the one discussed in Sec.~\ref{subsec:symmetry},
where one can change to new varibales in which $S_E$ becomes real
\cite{Caldeira,Ao}. Although we will often refer to the action functional 
of the form (\ref{euclid_action}), much of the results below apply
also to a more general retarded Euclidean action, which is of interest
for systems coupled to a bath.

In the spirit of the WKB approximation, the path integral
(\ref{stsum}) will be evaluated by the steepest descent method. The
extremal paths ${\bf r}(\tau)$ of the functional $S_E$ satisfy the
equation
\begin{equation}
\label{eom_b}
m{d^2{\bf r} \over d\tau^2}=\nabla U({\bf r})+i{e\over c}\left[ {d
{\bf r} \over d\tau}\times {\bf B}\right].
\end{equation}
The equation of motion (\ref{eom_b}) has to be solved with the
periodic boundary condition ${\bf r}(0)={\bf r}(\beta)$. Note that the
sign of the potential has been inverted compared to the case of
classical motion in real time.

For low temperatures and for the potential $U({\bf r})$, which is
parabolic near its intrawell minimum ${\bf r}_{\rm well}$,
Eq.~(\ref{eom_b}) has a solution ${\bf r}(\tau)={\bf r}_{\rm well}$,
with $S_E=0$. It gives the real part of the partition function, see
below.
%
As in the case $B=0$, the imaginary part of $Z$ is determined by
another solution of (\ref{eom_b}), which is of the bounce type. This
solution, ${\bf r}_{\rm b}(\tau)$, starts near ${\bf r}_{\rm well}$,
slides downhill in the inverted potential $-U({\bf r})$, and in time
$\beta$ comes back. For $B=0$ the corresponding path is a symmetric
real trajectory, ${\bf r}_{\rm b}(\tau)= {\bf r}_{\rm b}(\beta -
\tau)$, which bounces off the turning point $\dot{\bf r}_{\rm
b}(\beta/2) = 0$.

For $B\neq 0$, because of broken time-reversal symmetry, the path
${\bf r}_{\rm b}(\tau)$ is complex, and the velocity along this path
does not become equal to zero. The path is not symmetrical in time,
because the replacement $\tau \to -\tau$ changes
Eq.~(\ref{eom_b}). However, if we simultaneously 
change $i\to -i$, the equation remains unchanged. Therefore the
bounce-type path has the symmetry
\begin{equation}
\label{time-revers}
 {\bf r}_{\rm b}(\tau)={\bf r}_{\rm b}^*(\beta-\tau).
\end{equation}
An immediate and very important consequence of Eq.~(\ref{time-revers})
is that the value of $S_E({\bf r}_{\rm b})$ for the bounce-type path
is {\it real}. This value gives the tunneling exponent, see below.

\subsection{The eigenvalue problem}

The prefactor in $Z$ can be found by integrating over the tubes of
paths around the extremal paths. It can be done by expanding $S_E$ in
deviations from the extremal paths to the second order, and then expanding
${\bf r}(\tau)-{\bf r}_{\rm well}$ and ${\bf r}(\tau)-{\bf r}_{\rm
b}(\tau)$ in the eigenfunctions $\bfpsi_n(\tau)$ of the appropriate
eigenvalue problem,
\begin{eqnarray}
\label{d2s}
\hat{\bf\cal F}\bfpsi_n \equiv
\int\nolimits_0^{\beta} d\tau'\hat {\bf F}(\tau,\tau')\bfpsi_n(\tau') = 
\lambda_n \bfpsi_n(\tau), \nonumber  \\
 \hat F_{ij}(\tau,\tau')= 
\delta^2 S_E/ \delta  r_i(\tau) \delta  r_j(\tau').
%
\end{eqnarray}
Here, the derivatives of the action are calculated on the
corresponding extremal trajectory ${\bf r}_{\rm well}$ and ${\bf
r}_{\rm b}(\tau)$, and periodic boundary conditions are
assumed. The operator $\hat F$ is simplified for a non-retarded
action (\ref{euclid_action}), $\hat F_{ij}(\tau,\tau')
=\delta(\tau-\tau')\,\hat{\rm f}_{ij}(\tau)$.

For $B=0$, the operator $\hat {\bf \cal F}$ is Hermitian, with
$\hat{\rm f}_{ij} = -m \delta_{ij}(d^2 / d\tau^2) + \partial^2 U
/\partial r_i \partial r_j$, if the action is given by
(\ref{euclid_action}). Therefore the functions $\bfpsi_n(\tau)$ form
complete and orthogonal sets for each extreme trajectory
(\ref{eom_b}), and the eigenvalues $\lambda_n$ are real.

For $B\neq 0$, the operator $\hat {\bf\cal F}$ becomes non-Hermitian. For
example, in the case of a uniform magnetic field in
(\ref{euclid_action}), $\hat{\rm f}_{kl}(\tau)$ acquires an extra
term $i(e/c) \epsilon_{klj} B_j (d/ d\tau)$ ($\epsilon_{klj}$ is the
Levi-Civita symbol). Therefore some of the eigenvalues $\lambda_n$
become complex. The eigenvectors $\bfpsi_n$ with different $n$ are
orthogonal not to each other, but
to the eigenvectors $
\mbox{\boldmath $\phi$}_n $ of the Hermitian conjugate operator,
\begin{eqnarray*}
\int_0^{\beta} d\tau'\,\hat {\bf F}^\dagger(\tau,\tau')
\mbox{\boldmath $\phi$}_n(\tau')=\lambda_n^* \mbox{\boldmath $\phi$}_n(\tau).
\end{eqnarray*}
Taking into account the symmetry (\ref{time-revers}) of the
extremal trajectories, we find that
\begin{equation}
\label{F-symm}
\hat {\bf F}^\dagger(\tau,\tau')= 
\hat {\bf F}(\beta-\tau,\beta-\tau')=\hat{\bf F}^*(\tau,\tau').
\end{equation}
The energy spectra for several complex Hamiltonians with the
symmetry, which is similar to (\ref{F-symm}) (and was called the
${\cal PT}$-symmetry), were investigated earlier numerically and using
the WKB approximation\cite{Bender}.

The symmetry (\ref{F-symm}) has
several consequences. First, it shows that
$\bfpsi_n(\tau)=\alpha_n\bfphi_n^*(\tau)$, where $\alpha_n$ is a
constant.  This means that, with proper normalization, the
orthogonality relation becomes
\begin{equation}
\label{ortho}
\int\nolimits_0^{\beta}d\tau\,
\bfpsi_m(\tau)\bfpsi_n(\tau)=\delta_{mn}
\end{equation}
(here, we assumed that the eigenvalues are nondegenerate; for degenerate
eigenvalues, the condition (\ref{ortho}) can be satisfied by choosing
appropriate linear combinations of the eigenfunctions with same $\lambda_n$).

It also follows from Eq.~(\ref{F-symm}) that, if $\bfpsi_n(\tau)$ is
an eigenfunction of (\ref{d2s}) with an eigenvalue $\lambda_n$, then
$\bfpsi^*_n(\beta - \tau)$ is also an eigenfunction of the same
boundary value problem, but with the eigenvalue $\lambda_n^*$. This
means that a part of the eigenvalues $\lambda_n$ in (\ref{d2s}) are
real, whereas another part are pairs of complex conjugate numbers.

Pairs of complex conjugate eigenvalues are formed in the following
way. For $B=0$, all eigenvalues are real. With increasing $B$ some
eigenvalues approach each other, pairwise, while still remaining
real. Eventually they merge, and for larger $B$ become complex
conjugate, as described in Appendix~\ref{app2}. For 1D
Schr\"odinger-type equation with complex Hamiltonians such behavior with
the varying control parameter was indeed observed numerically
\cite{Bender}.

\subsubsection{Eigenvalues near the potential well}

As an illustration, we consider the eigenvalue problem near ${\bf
r}_{\rm well}$ for the action functional (\ref{euclid_action}). Here,
Eq.~(\ref{d2s}) becomes linear, and the eigenfunctions
$\bfpsi_n(\tau)$ can be sought in the form of linear combinations of
$\exp(\pm i\omega_n\tau)$, with $\omega_n=2\pi n/\beta$. 
The eigenvalues are obtained from the equation
\begin{equation}
\label{near_well}
\det \left[(m\omega_n^2 - \lambda_{n\,\nu})\delta_{kl} + m\Omega^2_{kl} - 
{e\over c}\omega_n\epsilon_{klj}B_j\right] = 0,
\end{equation}
where $m\Omega^2_{kl}=[\partial^2 U/\partial r_k\partial r_l]_{{\bf
r}_{\rm well}}$, and ${\bf B}$ is the magnetic field at ${\bf r}_{\rm
well}$. The subscript $\nu$ enumerates the eigenvalues $\lambda$ for a
given Matsubara frequency.

If, for example, ${\bf B}$
is pointing along a principal axes of the tensor $\Omega^2_{kl}$ (say,
the axes 1), then we have $\lambda_{n\,1}/m = \omega_n^2 +
\Omega_1^2$, and
\begin{eqnarray}
\label{conjugate_eigen}
m^{-1}\lambda_{n\,2,3}=&&
\omega_n^2 + {1\over 2}(\Omega_2^2+ \Omega_3^2) \nonumber \\ 
&&\pm {1\over 2}
\left[(\Omega_2^2-\Omega_3^2)^2 - 4\omega_c^2\omega_n^2\right]^{1/2}
\end{eqnarray}
where $\Omega^2_{\nu}> 0$ are the principal values of the tensor
$\Omega^2_{kl}$.  Clearly, the eigenvalues $\lambda_{n\,2,3}$ are
complex conjugate pairs, for large enough $\omega_n^2\omega_c^2$.

Eq.~(\ref{conjugate_eigen}) shows explicitly also how pairs of complex
eigenvalues emerge with varying magnetic field as a result of merging
of adjacent real eigenvalues, as discussed for the general case in
Appendix~\ref{app2}.

\subsubsection{Eigenvalues for the bounce trajectory}

A specific feature of the eigenvalue problem (\ref{d2s}) for the
bounce trajectory ${\bf r}_b(\tau)$ at low temperatures is that one
of the eigenvalues is $\lambda_1=0$. It corresponds to the
eigenfunction $\bfpsi_1(\tau)\propto \dot{\bf r}_{\rm b}(\tau)$. 
%
For $B=0$, the vector function $\bfpsi_1(\tau)\propto\dot{\bf r}_{\rm
b}(\tau)$ has one zero for all components. Therefore, as can be shown
using standard arguments, it is the eigenfunction of the first excited
state of the multicomponent Schr\"odinger-type equation (\ref{d2s})
(except for a nongeneric case where the components of $\bfpsi$
separate). Since the eigenvalue problem (\ref{d2s}) is Hermitian for
$B=0$, all eigenvalues $\lambda_n$ with $n\geq 2$ are positive, and
the eigenvalue of the ground state is negative,
$\lambda_0<0$. \cite{Langer}.

We are not aware of the oscillation theorem for non-Hermitian
problems. However, since $\bfpsi_1(\tau)\propto \dot{\bf r}_{\rm
b}(\tau)$ is an eigenfunction with zero eigenvalue even for $B\neq 0$,
as $B$ increases from zero, the eigenvalue $\lambda_1$ does not merge
with other real eigenvalues to form a pair of complex conjugate
eigenvalues. Therefore, pairs of complex
conjugate eigenvalues will be only formed from $\lambda_n$ that were
positive for $B=0$. The negative root $\lambda_0$ will remain real and
negative. In principle, as a result of coalescence of complex conjugate
eigenvalues, there may emerge pairs of negative real
eigenvalues. However, the total number of negative real eigenvalues
will be odd.

\subsection{The prefactor}

We are now in a position to calculate the prefactor in the partition
function $Z$. The standard step is to expand the deviation $\delta
{\bf r}(\tau)$ of the integration path in (\ref{stsum}) from the
extreme trajectory ${\bf r}_{\rm well}$ or ${\bf r}_{\rm b}$ in terms
of the  eigenfunctions $\bfpsi_n$ of the corresponding eigenvalue problem,
$\delta
{\bf r}(\tau)=\sum c_n
\bfpsi_n(\tau)$.
With account taken of the orthogonality condition (\ref{ortho}), the
increment $\delta S_E$ of the Euclidean action related to the
deviation of the trajectory $\delta {\bf r}$ then becomes $\delta S_E
= \sum \lambda_nc_n^2/2$.

The above expansion assumes that the set $\{ \bfpsi_n\}$ is
complete. The completeness is known for $B=0$, where the eigenvalue
problem (\ref{d2s}) is Hermitian. As $B$ changes, the number of states
does not change. From the orthogonality condition (\ref{ortho}), none
of the wave functions becomes a linear combination of other wave
functions. This makes us believe that the functions $\bfpsi_n$ form a
complete set even for $B\neq 0$ and justifies the above expansion.

The path integral (\ref{stsum}) can be obtained as a limit $N\to
\infty$ of integrals over $d{\bf r}(\tau_k)$ at discretized instants of
time $\tau_k=k\Delta\tau, \, \Delta\tau = \beta/N$. In the standard
way, we change to integration over $dc_n$. Because of the
orthogonality relation (\ref{ortho}), the determinant
$\det[\bfpsi_n(\tau_k)]$ of the transformation of variables is real
and is equal to $\pm (\Delta \tau)^{N/2}$. Integration of
$\exp(-\delta S_E)$ over $dc_n$ gives ${\rm const}\times
\prod_n\lambda_n^{-1/2}$. 

Let us now consider the contribution to the partition function $Z_{\rm
well}$ from trajectories close to the potential minimum ${\bf r}_{\rm
well}$. The corresponding eigenvalues $\lambda_n^{\rm (well)}$ are
either positive or belong to complex conjugate pairs,
cf. ~(\ref{conjugate_eigen}). Therefore $Z_{\rm well}$ is real.  Since
$S_E[{\bf r}_{\rm well}]=0$, there is no exponentially small factor in
$Z_{\rm well}$. This term gives the partition function for low-lying
intrawell excitations in the presence of the magnetic field.

In evaluating the contribution $Z_{\rm b}$ from paths close to the
bounce trajectory, special care has to be taken of the eigenvalue
$\lambda_1^{\rm (b)} = 0$. A standard analysis \cite{Langer,Coleman}
shows that integration over $dc_1$ gives the factor $\beta$ in $Z_{\rm
b}$. The positive and complex conjugate eigenvalues $\lambda_n^{\rm (b)}$
give a real positive factor in $Z_{\rm b}$, whereas the negative
eigenvalue $\lambda_0^{\rm (b)}$ (or an odd number of negative eigenvalues) make
$Z_{\rm b}$ purely imaginary. In addition, $Z_{\rm b}$ contains the
exponential factor $\exp\{-S_E[{\bf r}_{\rm b}(\tau)]\}$. Overall,
this gives the tunneling rate (\ref{rate})
\begin{equation}
\label{final_rate}
W\approx 2T|Z_{\rm b}|/Z_{\rm well}\propto \exp\{-S_E[{\bf r}_{\rm
b}(\tau)]\}. 
\end{equation}

Eq.~(\ref{final_rate}) shows that the instanton technique can be used
in the presence of a magnetic field in spite of the fact that the
field breaks time-reversal symmetry. The actual calculation is in many
respects different from that for $B=0$. In particular, the bounce
trajectory is complex.  The eigenvalues which determine the prefactor
should be found from a non-Hermitian eigenvalue problem. They may be
complex, in which case they form pairs of complex-conjugate numbers.
The bounce trajectory touches the real escape trajectory at a real
point ${\bf r}_{\rm b}(\beta/2)$, with a finite real velocity $\dot
{\bf r}_{\rm b}(\beta/2)$. However, from our general WKB analysis of
tunneling, it follows that this is not the point where the particle
``shows up'' as a semiclassical wave packet.

\section{conclusions}
\label{sec:conc}

In conclusion, the problem of tunneling in a magnetic
field can be solved in the semiclassical limit by analyzing the
Hamiltonian trajectories of the particle in complex phase space and
time. The boundary conditions are determined by the intrawell wave
function and its analytic continuation. This approach allows one to
find both the tunneling exponent and the tail of the wave function of
the localized state.  It does not require to consider either a part of
the potential or the magnetic field as a perturbation, and it
can be applied to a three-dimensional potential of a general form.

The escape rate is generally {\it exponentially} smaller than the
probability for a particle to reach the boundary of the classically
accessible range $U({\bf r})= E$.  The escaped particle ``shows up''
from the tunneling barrier with finite velocity and behind the surface
$U({\bf r})= E$. The connection of the decaying and propagating waves
occurs on caustics of the set of the Hamiltonian trajectories, where
the action is branching. The caustics are complex surfaces in 3D
space. In the presence of a magnetic field, they go through real space
along lines (instead of surfaces, for $B=0$).

An interesting feature of tunneling in a magnetic field is the
occurrence of a switching surface, where there merge different WKB
branches of the wave function. The slope of the {\it logarithm} of the
wave function sharply changes at the switching surface, from its value
on one of the branches to that on the other branch. The escaped
particle first shows up as a propagating semiclassical wave packet on
the switching surface. It happens where the classical escape
trajectory crosses the switching surface. 

The switching surfaces are observable via experiments in which the
electron density is measured, although such experiments are extremely
hard to do. For tunneling from 2D electron systems in
heterostructures, one can think of NMR experiments with samples that
contain delta-doped layers of nuclear spins (of an isotope that
differs from that in the host material). If the nuclei are
sufficiently far from the electron layer, they will detect the local
electron density on the tail of the wave function and its variation
with varying fields. One could also use light-scattering measurements.

Switching between branches of the WKB wave function for $B\neq 0$ is
similar to the switching between different branches of the probability
distribution in classical systems away from thermal equilibrium. Such
systems lack time-reversal symmetry, as do also quantum systems in a
magnetic field. The tail of the classical distribution is formed by
infrequent fluctuations. Fluctuational paths to a given state from the
equilibrium position (attractor) form a narrow tube centered at the
most probable path. This path is given by a solution of the variational
problem of finding the maximum of the logarithm of the probability
distribution \cite{Freidlin:84,DK:79,Graham:84}. In many cases of
physical interest the corresponding Euler equations are similar to
Eqs.~(\ref{eom}).  However, in contrast to the tunneling problem,
classical optimal paths can be observed\cite{prehistory}.  Switching
surfaces in the phase space of fluctuating nonequilibrium systems
separate areas reached along topologically different optimal
paths\cite{DMS}$^{\rm (a)}$. They have been seen in analog simulations
\cite{DMS}$^{\rm (b)}$.

It follows from the results of this paper that, for potential wells
which are parabolic near the minimum, even in the presence of a
magnetic field one can still use the instanton technique in order to
find the escape rate. However, the bounce trajectory, which gives the
tunneling exponent, is now complex. Also in contrast to the $B=0$
case, the evaluation of the prefactor requires solving a non-Hermitian
boundary value problem, which generally has pairs of complex conjugate
eigenvalues.

The results for the model of tunneling from a strongly correlated 2D
electron system illustrate the general conclusions about tunneling in
a magnetic field. They show that the developed method allows us to
find the tunneling rate and the wave function in a generic system
which does not have any special symmetry. They also confirm the
general conclusions about the structure of singularities related to
the branching of the WKB wave function, and the occurrence of the
switching line. We found that the tunneling rate in the magnetic field
is highly sensitive to the in-plane electron dynamics and
exponentially increases when electrons are more strongly confined in
the plane. It also increases if electrons in the 2D layer can
adjust to the tunneling electron and thus decrease the potential
barrier.

This research was supported in part by the NSF through
Grant No. PHY-0071059.

\appendix

\section{The many-electron Hamiltonian}
\label{app1}

A simple and important model which allows us to consider the effect of
electron correlations on tunneling from a 2D electron system is the
model of a Wigner crystal. In this model, the in-plane electron motion
is small-amplitude vibrations about equilibrium positions. Because of
strong correlations, exchange effects are not important, and the
tunneling electron can be identified. Its tunneling motion is affected
by the interaction with other electrons. 

We will assume that the equilibrium in-plane position of the tunneling
electron is at the origin. Then, in the presence of a magnetic field ${\bf
B}$ parallel to the electron layer, the full Hamiltonian is of the
form
\begin{equation}
\label{wigner_hamiltonian}
H={p_z^2\over 2m} +U_0(z) +H_v +H_B,
\end{equation} 
with
\begin{equation}
\label{Hv}
H_v= {1\over 2}\sum_{{\bf k},j}\left[m^{-1}{\bf p}_{{\bf k}j}{\bf
p}_{-{\bf k}j} + m \omega_{{\bf k}j}^2 {\bf u}_{{\bf k}j}{\bf
u}_{-{\bf k}j}\right] 
\end{equation}
and
\begin{eqnarray}
\label{interaction}
H_B= && {1 \over 2} m\omega_c^2 z^2 - \omega_c z N^{-1/2}
\sum_{{\bf k},j}[\hat{\bf B}\times{\bf p}_{{\bf k}j}]_z \nonumber\\
&& + U_{\rm int}(z,\{{\bf u}_{{\bf k}j}\}).
\end{eqnarray}
Here, ${\bf p}_{{\bf k}j}$, ${\bf u}_{{\bf k}j}$, and $\omega_{{\bf
k}j}$ are the 2D momentum, displacement, and frequency of the WC
phonon of branch $j$ ($j=1,2$) with a 2D wave vector ${\bf k}$. The
in-plane momentum of the tunneling electron is $N^{-1/2}\sum{\bf
p}_{{\bf k}j} $ for $B=0$. The term $U_0(z)$ describes the tunneling
barrier [cf. Eq.~(\ref{U_0(z)})] for the electron at the origin
provided all other electrons are at their in-plane lattice sites.

The term $H_B$ couples the out-of-plane tunneling motion to lattice
vibrations. The problem of many-electron tunneling is thus mapped onto
a familiar problem of a particle coupled to a bath of harmonic
oscillators
\cite{Caldeira,Feynman}. A part of the coupling is due purely to the
magnetic field. Another part comes from the term $U_{\rm int}$, which
describes the change of the tunneling barrier because of  electron
vibrations. Its simplest form is given by the lowest-order term of
the expansion of the electron energy,
\begin{eqnarray}
\label{Uint}
U_{\rm int}(z,\{{\bf u}_{{\bf k}j}\}) = z\sum\nolimits_{{\bf k},j}
{\bf g}_{-{\bf k}j}{\bf u}_{{\bf k}j}
\end{eqnarray}
where ${\bf g}_{{\bf k}j}$ are coupling constants [for
electrons on a thick helium film \cite{AP} the major term in $U_{\rm int}$ is
$\propto z^2\,$]. The coupling (\ref{Uint}) leads to
lowering of the tunneling barrier as a result of appropriate
displacements of the electrons surrounding the tunneling electron.

The major effect on tunneling comes from high-frequency in-plane
vibrations, which have large density of states
\cite{Dykman-01}. Therefore it is not unreasonable to 
use the Einstein model of the Wigner crystal, in which all vibrations
have same frequency $\omega_0$. Then, except for the term $U_{\rm
int}$, the Hamiltonian (\ref{wigner_hamiltonian}) becomes a sum of
Hamiltonians of  confined in the plane noninteracting electrons. The
Hamiltonian of the tunneling electron has the form
(\ref{hamiltonian}), with the potential $U({\bf r})$ given by
(\ref{U(z)}).

For $B=0$, the out-of-plane motion of the tunneling electron is
coupled only to in-plane displacements of {\it other} electrons. In
the Einstein model, it means that the out-of-plane motion is decoupled
from the in-plane motion of the tunneling electron itself. Instead, it
is coupled to an in-plane oscillator with the coordinate given by a
(totally symmetric) linear combination of displacements of the other
electrons. This maps the problem onto the problem discussed in
Sec.~\ref{sec:asymmetry}, with the in-plane electron coordinate $x$ in
Eq.~(\ref{U(x,z)}) corresponding to the coordinate of this oscillator,
and with $\mu$ being a linear combination of the (weighted)
coefficients ${\bf g}_{{\bf k}j}$. Of course, for $B\neq 0$ this
mapping no longer applies, and extra degrees of freedom have to be taken
into account. Yet we expect that the model (\ref{U(x,z)}) catches
much of the qualitative features of many-electron tunneling.

\protect\section{Emergence of complex eigenvalues}
\label{app2}

In this appendix we consider how, with the varying control parameter,
two real eigenvalues merge and then become complex. Near this
transition, the eigenvalues can be sought by perturbation theory. We
start with a value of ${\bf B}={\bf B}_0$ (we can also use another
control parameter), where the given adjacent eigenvalues $\lambda_n,\,
\lambda_m$ are close to each other and are real. For small $|\delta
{\bf B}|=|{\bf B}-{\bf B}_0|$, the functional $\hat{\bf\cal F}$ is
close to its value for ${\bf B}_0$, $
\hat{\bf\cal F} \approx 
\hat{\bf\cal F}_0 + \delta \hat{\bf\cal F}$ in (\ref{d2s}). 
To first order in $\delta \hat{\bf\cal F}$, the eigenvalues are given
by the expressions $[\lambda_n({\bf B}_0) + \lambda_m({\bf B}_0)]/2 +
\lambda_{\pm}$, with
\wid
\[
\hfill \lambda_{\pm}= {1\over 2}
\left(\delta\hat{\cal F}^{nn} + \delta\hat{\cal F}^{mm}\right)
\pm {1\over 2}\left[
\left(\delta\hat{\cal F}^{nn} - \delta\hat{\cal F}^{mm}-\delta\lambda\right)^2 
+4\delta\hat{\cal F}^{nm}\delta\hat{\cal F}^{mn}\right]^{1/2},
\; 
\delta\hat{\cal F}^{nm}=\langle \bfpsi_n|\delta\hat{\bf\cal F}|
\bfpsi_m\rangle. \hfill ({\rm B}1)
\]
\nar
\noindent
with the wave functions calculated for ${\bf B}_0$, and with
$\delta\lambda = \lambda_m-\lambda_n$ for ${\bf B=B}_0$.

Because of the symmetry (\ref{F-symm}), the matrix elements of
$\delta\hat{\bf\cal F}$ in (B1) are real. However, the
product $\delta\hat{\cal F}^{nm}\delta\hat{\cal F}^{mn}$ does not have
to be positive, and in fact we are interested in the case where it is
negative. In this case, instead of level anticrossing, we have the
dependence of the eigenvalues on the distance $\delta \lambda$ shown
in Fig.~11. In the gap the eigenvalues are complex
conjugate. 

\begin{figure}
\begin{center}
\epsfxsize=3.3in                
\leavevmode\epsfbox{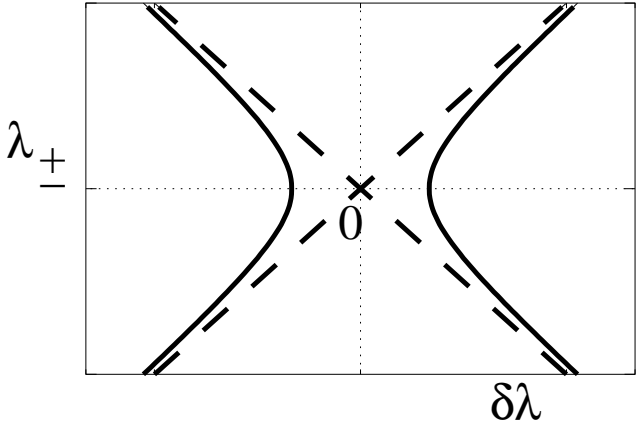}
\end{center}
%
%
\caption{The dependence of the shifted eigenvalues $\lambda_{\pm}$ on 
the distance $\delta\lambda$ between the eigenvalues for ${\bf B=B}_0$
(schematically). We count $\delta\lambda$ off from $\delta\hat{\cal
F}^{nn}-\delta\hat{\cal F}^{mm}$, and $\lambda_{\pm}$ from  $(\delta\hat{\cal
F}^{nn}+\delta\hat{\cal F}^{mm})/2$.}
\end{figure}

We note that the control parameter in the physical system is not
$\delta \lambda$, and it may be more interesting to look at the
eigenvalues as functions of the amplitude of $\delta\hat{\cal
F}^{nm}$. Their behavior is similar to what is shown in Fig.~11, if
the diagonal and off-diagonal matrix elements depend on the control
parameter in the same way. Otherwise, once the eigenvalues become
complex with changing control parameter, they do not have to become
real again, as is the case for the eigenvalues given by
Eqs.(\ref{conjugate_eigen}) as functions of $\omega_c$. We note that
there is also an opposite process of merging of complex conjugate
eigenvalues, which is also described by Eq.~(B1).

\end{multicols}

\begin{thebibliography}{99} 

\bibitem[*]{byline} To whom correspondence should be addressed,
e-mail: dykman@pa.msu.edu

\bibitem{Efros-Shklovskii} B.I. Shklovskii and A.L. Efros, 
{\it Electronic Properties of Doped Semiconductors} (Springer-Verlag, NY 1984).

\bibitem{DasSarma_book-97} For reviews see J.P. Eisenstein, 
in ``Perspectives in Quantum Hall Effects'', ed. by S. Das Sarma and
A. Pinczuk (Wiley, NY 1997), p.~37; C.L. Kane and M.P.A.~Fisher, {\it
ibid.}, p.~109, S.M. Girvin and A.H.~MacDonald, {\it ibid.}, p.~161;
B.I.~Halperin, {\it ibid.} p.~225.

\bibitem{Ashoori} N.B.~Zhitenev, M. Brodsky, and R.C.~Ashoori, M.R.~Melloch
Phys. Rev. Lett. {\bf 77}, 1833 (1996); M.B.~Hastings and
L.S.~Levitov, Phys. Rev. Lett. {\bf 77}, 4422 (1996).

\bibitem{Chang} M. Grayson, D.C. Tsui, L.N. Pfeiffer, K.W. West, and 
A.M. Chang, Phys. Rev. Lett. {\bf 86}, 2645 (2001) and references
therein.

\bibitem{Eisenstein-2000}I.B.~Spielman, J.P.~Eisenstein, 
L.N. Pfeiffer, and K.W. West, Phys. Rev. Lett. {\bf 84}, 5808 (2000).

\bibitem{Smoliner} J. Smoliner, W.~Demmerle, G.~Berthold, E.~Gornik,
G.~Weimann, and W. Schlapp, Phys. Rev. Lett. {\bf 63}, 2116 (1989); G. Rainer,
J.~Smoliner, E.~Gornik, G. B\"ohm, and G. Weimann, Phys. Rev. B {\bf
51}, 17642 (1995).

\bibitem{Eisenstein} J.P.~Eisenstein, T.J.~Gramila, L.N.~Pfeiffer, and
K.W. ~West, Phys.  Rev. B {\bf 44}, 6511 (1991); S.Q.~Murphy,
J.P.~Eisenstein, L.N.~Pfeiffer, and K.W.~West, Phys.  Rev. B {\bf 52},
14825 (1995).

\bibitem{MacDonald} L. Zheng and A.H. MacDonald, 
Phys. Rev. B {\bf 47}, 10619 (1993).

\bibitem{minigaps} T.~Ihn, H.~Carmona, P.C.~Main, L.~Eaves, and
M.~Henini, Phys. Rev. B {\bf 54}, R2315 (1996); M.J.~Yang,
C.H.~Yang, B.R.~Bennett, and B.V.~Shanabrook, Phys. Rev. Lett. {\bf
78}, 4613 (1997); M.~Lakrimi, S.~Khym, R.J.~Nicholas,
D.M.~Symons, F.M.~Peeters, N.J.~Mason, and P.J.~Walker,
Phys. Rev. Lett. {\bf 79}, 3034 (1997).

\bibitem{Andreimagn} L. Menna, S. Y\"{u}cel, and E.Y.~Andrei, 
Phys. Rev. Lett.  {\bf 70}, 2154 (1993); S. Y\"{u}cel, L. Menna,
and E.Y. Andrei, Physica B {\bf 194 -- 196}, 1223 (1994).

\bibitem{Dykman-01}M.I.~Dykman, T.~Sharpee, and P.M.~Platzman, 
Phys. Rev. Lett {\bf 86}, 2408 (2001); T.~Sharpee, M.I.~Dykman,
P.M.~Platzman, cond-mat/0103151.

\bibitem{Abrahams-00} E. Abrahams, S.V. Kravchenko, and M.P. Sarachik, 
Rev. Mod. Phys. {\bf 73}, 251 (2001).

\bibitem{Popov} (a) L.P. Kotova, A.M. Perelomov, and V.S. Popov, Sov. Phys. 
JETP {\bf 27}, 616 (1968); (b) V.S. Popov, B.M. Karnakov, and V.D. Mur,
Sov. Phys. JETP {\bf 86}, 860 (1998).

\bibitem{Caldeira} A.O. Caldeira and A.J. Leggett,  Ann. Phys. {\bf 149}, 
374 (1983).

\bibitem{Fert_Halperin} H.A.~Fertig and B.I.~Halperin, Phys. Rev. B
{\bf 36}, 7969 (1987).

\bibitem{Ao} P. Ao,  Phys. Rev. Lett. {\bf 72}, 1898
(1994); Phys. Scripta {\bf T69}, 7 (1997).

\bibitem{Shklovskii} B.I. Shklovskii,  JETP Lett. {\bf 36},
51 (1982).

\bibitem{Thouless} Qin Li and D.J.~Thouless,  Phys. Rev. B {\bf 40},
9738 (1989).

\bibitem{Feng} T.~Martin and S.~Feng, Phys. Rev. B {\bf 44}, 9084 (1991).

\bibitem{Hajdu} J.~Hajdu, M.E.~Raikh, and T.V.~Shahbazyan,
Phys. Rev. B {\bf 50}, 17625 (1994); M.E.~Raikh and
T.V.~Shahbazyan, Phys. Rev. B {\bf 51}, 9682 (1995)

\bibitem{Helffer} B.~Helffer and J.~Sj\"{o}strand,  Ann. Scuola Norm. Sup. Pisa Cl. Sci. (4) {\bf 14},
625 (1988).


\bibitem{Nakamura} S.~Nakamura, Comm. Math. Phys. {\bf 200}, 25 (1999)
and references therein.

\bibitem{Barabash-2000}T. Barabash-Sharpee, M.I. Dykman, P.M. Platzman, Phys. Rev. Lett. {\bf 84}, 2227 (2000)

\bibitem{Langer} J.S. Langer, Ann. Phys. {\bf 41}, 108 (1967).

\bibitem{Coleman} S.~Coleman,  Phys. Rev. D {\bf 15}, 2929 (1977);
C.G.~Callan and S.~Coleman, Phys. Rev. D {\bf 16}, 1762 (1977).

\bibitem{AK} A. Auerbach and S. Kivelson, Nucl. Phys. {\bf B257}, 799 (1985).

\bibitem{Schmid} U. Eckern and A. Schmid, in {\it Quantum Tunnelling
in Condensed Matter}, eds. Yu. Kagan and A.J. Leggett (Elsevier, NY
1992), p.~145.

\bibitem{Huang} Z.H.~Huang, T.E.~Feuchtwang, P.H.~Cutler, and
E.~Kazes, Phys. Rev. A {\bf 41}, 32 (1990).

\bibitem{Knoll} J. Knoll and R. Schaeffer, Ann. Phys. {\bf 97}, 307 (1976).

\bibitem{Berry} M.V. Berry, Adv. Phys. {\bf 25}, 1 (1976);
Proc. R. Soc. Lond. A {\bf 422}, 7 (1989);
Proc. R. Soc. Lond. A {\bf 427}, 265 (1990); J. Heading, {\it
An Introduction to Phase-Integrals Methods} (London: Methuen, 1962).

\bibitem{Schulman_book} L.S. Schulman, 
{\it Techniques and applications of path integration} (Wiley, New
York, 1981).

\bibitem{LandauQM} L.D. Landau and E.M. Lifshitz, {\it Quantum
mechanics: non-relativistic theory} (Pergamon, NY 1977).

\bibitem{Stokes}G.G. Stokes,  Trans. Camb. Phil. Soc., 
{\bf 10}, 106 (1857).

\bibitem{Berry-Mount}M.V. Berry and K.E. Mount, Rep. Progr. 
Phys. {\bf 35}, 315 (1972).

\bibitem{AP} M.Ya. Azbel, Phys. Rev. Lett. {\bf 64}, 1553 (1990);
M.Ya. Azbel and P.M. Platzman, Phys. Rev. Lett. {\bf 65}, 1376 (1990).

\bibitem{Feynman}
R. P. Feynman and A. R. Hibbs, {\em Quantum Mechanics and Path Integrals}
(McGraw-Hill, New York, 1965).

\bibitem{Bender} C.M.~Bender and S.~Boettcher, Phys. Rev. Lett. {\bf
80}, 5243 (1998); C.M.~Bender, S.~Boettcher, P.N.~Meisinger,
J. Math. Phys {\bf 40}, 2201 (1999); C.M.~Bender, M.~Berry,
P.~N.~Meisinger, van~M~ Savage, and M.~Simsek, J. Phys. A {\bf 34},
L31 (2001).

\bibitem{Freidlin:84}
M.I. Freidlin and A.D. Wentzel, {\em Random Perturbations in Dynamical
Systems} (Springer, New-York, 1984).

\bibitem{DK:79}
M.I. Dykman and M.A. Krivoglaz, Sov. Phys. JETP {\bf 50},
30 (1979); in {\em Soviet Physics Reviews}, edited by I.M.
Khalatnikov (Harwood Academic Publishers, New York, 1984), Vol.~5, 
265; M.I.~Dykman Phys. Rev. A {\bf 42}, 2020 (1990).

\bibitem{Graham:84} R. Graham and T. T\'{e}l, Phys. Rev. Lett. {\bf
52}, 9 (1984); R. Graham and T. T\'{e}l, Phys. Rev. A {\bf 31}, 1109
(1985); R. Graham, in {\em Noise in Nonlinear Dynamical Systems},
edited by F. Moss and P.V.~E. McClintock (Cambridge University Press,
Cambridge, 1989), Vol.~1, 225.

\bibitem{prehistory} M.I.~Dykman, P.V.E.~McClintock, V.N.~Smelyanskiy, 
N.D.~Stein, and N.G. Stocks, Phys. Rev. Lett. {\bf 68}, 2718 (1992).

\bibitem{DMS} (a) M.I. Dykman, M.M. Millonas, and V.N. Smelyanskiy,
Phys. Lett. A {\bf 195}, 53 (1994); (b) M.I. Dykman, D.G. Luchinsky,
P.V.E. McClintock, and V.N. Smelyanskiy, Phys. Rev. Lett. {\bf 77},
5229 (1996).


\end{thebibliography}
\end{document}